\documentclass[fleqn,usenatbib]{mnras}

\usepackage{newtxtext,newtxmath}

\usepackage[T1]{fontenc}

\DeclareRobustCommand{\VAN}[3]{#2}
\let\VANthebibliography\thebibliography
\def\thebibliography{\DeclareRobustCommand{\VAN}[3]{##3}\VANthebibliography}

\usepackage{graphicx}	
\usepackage{amsmath}	
\usepackage{times} 
\usepackage{latexsym}
\usepackage{amsfonts}
\usepackage{rotfloat} 
\usepackage{rotating} 
\usepackage{float}
\usepackage{natbib}
\usepackage{supertabular}
\usepackage{longtable}
\usepackage{url}
\usepackage{dcolumn}
\usepackage{placeins}
\usepackage{multirow} 
\usepackage{textcomp}
\usepackage{multicol}
\usepackage{bm}
\usepackage{pdflscape}
\usepackage{verbatim}
\usepackage{ae,aecompl}
\usepackage[utf8]{inputenc}
\usepackage[caption=false]{subfig}

\newcommand{\mgii}{\mbox{Mg\,{\sc ii}}}
\newcommand{\mgi}{\mbox{Mg\,{\sc i}}}
\newcommand{\feii}{\mbox{Fe\,{\sc ii}}}

\newcommand{\civ}{\mbox{C\,{\sc iv}}}

\newcommand{\sifour}{\mbox{Si\,{\sc iv}}}
\newcommand{\lya}{\ensuremath{{\rm Ly}\alpha}}

\newcommand{\zqso}{$z_{\rm QSO}$}
\newcommand{\zlen}{$z_{\rm lens}$}

\newcommand{\mhalo}{$\rm M_{\rm h}$}

\newcommand{\kms}{km\,s$^{-1}$}

\newcommand{\ergscm}{$\rm erg\,s^{-1}\,cm^{-2}$}

\newcommand{\msun}{$\rm M_\odot$}

\title[Metal absorption towards lensed quasars]{Probing coherence in metal absorption towards multiple images of strong gravitationally lensed quasars}

\author[R. Dutta et al.]{
Rajeshwari Dutta,$^{1,2,3}$\thanks{E-mail: rajeshwari.dutta@iucaa.in}
Ana Acebron,$^{4,5}$
Michele Fumagalli,$^{2,6}$
Claudio Grillo,$^{4,5}$
Gabriel B. Caminha,$^{7,8}$
\newauthor
Matteo Fossati$^{2,3}$
\\
$^{1}$ IUCAA, Postbag 4, Ganeshkind, Pune 411007, India \\
$^{2}$ Dipartimento di Fisica G. Occhialini, Universit\`a degli Studi di Milano Bicocca, Piazza della Scienza 3, 20126 Milano, Italy \\
$^{3}$ INAF - Osservatorio Astronomico di Brera, via Bianchi 46, 23087 Merate (LC), Italy \\
$^{4}$ Dipartimento di Fisica, Università degli Studi di Milano, Via Celoria 16, I-20133 Milano, Italy \\
$^{5}$ INAF - IASF Milano, via A. Corti 12, I-20133 Milano, Italy \\
$^{6}$ INAF – Osservatorio Astronomico di Trieste, via G. B. Tiepolo 11, I-34143 Trieste, Italy \\
$^{7}$ Technical University of Munich, TUM School of Natural Sciences, Department of Physics, James-Franck-Str 1, 85748 Garching, Germany \\
$^{8}$ Max-Planck-Institut f\"ur Astrophysik, Karl-Schwarzschild-Str. 1, D-85748 Garching, Germany \\
}

\date{Accepted XXX. Received YYY; in original form ZZZ}

\pubyear{2023}

\begin{document}
\label{firstpage}
\pagerange{\pageref{firstpage}--\pageref{lastpage}}
\maketitle

\begin{abstract}

We present a tomographic analysis of metal absorption lines arising from the circumgalactic medium (CGM) of galaxies at $z\approx$ 0.5--2, using Multi Unit Spectroscopic Explorer (MUSE) observations of two background quasars at $z\approx$ 2.2 and 2.8, which are two of the few currently known quasars with multiple images due to strong gravitational lensing by galaxy clusters at $z\approx$ 0.6 and 0.5, respectively. The angular separations between different pairs of quasar multiple images enable us to probe the absorption over transverse physical separations of $\approx$0.4--150 kpc, which are based on strong lensing models exploiting MUSE observations. The fractional difference in rest-frame equivalent width ($\Delta W_r$) of \mgii, \feii, \civ\ absorption increases on average with physical separation, indicating that the metal-enriched gaseous structures become less coherent with distance, with a likely coherence length scale of $\approx$10\,kpc. However, $\Delta W_r$ for all the ions vary considerably over $\approx$0.08-0.9, indicating a clumpy CGM over the full range of length scales probed. At the same time, paired \mgii\ absorption is detected across $\approx$100--150 kpc at similar line-of-sight velocities, which could be probing cool gas clouds within the same halo. No significant dependence of $\Delta W_r$ is found on the equivalent width and redshift of the absorbing gas and on the galaxy environment associated with the absorption. The high-ionization gas phase traced by \civ\ shows a higher degree of coherence than the low-ionization gas phase traced by \mgii, with $\approx$90 percent of \civ\ systems exhibiting $\Delta W_r$ $\le$0.5 at separations $\le$10 kpc compared to $\approx$50 percent of \mgii\ systems.

\end{abstract}




\section{Introduction}
\label{sec_introduction}

By regulating the flow of baryons in and out of galaxies, the circumgalactic medium (CGM) plays a vital role in galaxy evolution \citep{Tumlinson2017}. On one hand, pristine gas is accreted from the large-scale structures through the CGM onto the galaxy, while on the other hand, star formation and nuclear activity in the galaxy drive metal-enriched outflows into the CGM. The expelled metals either fall back onto the galaxy or spread in the intergalactic medium. Studying the distribution and mixing of metals around galaxies is therefore crucial to understanding the star formation history and chemical enrichment of galaxies \citep{Peroux2020}.

Due to its low density, the CGM has been most effectively probed in absorption against bright background sources such as quasars. Large surveys have statistically mapped the distribution of metal absorption lines around galaxies across different gas phases and redshifts \citep[e.g.][]{Tumlinson2011,Nielsen2013,Turner2014,Dutta2020,Dutta2021,Galbiati2023}. Ionization modeling of absorption line systems has been used to place constraints on the average physical conditions of the gas clouds such as metallicity, density, and size \citep[e.g.][]{Werk2014,Fumagalli2016,Haislmaier2021,Sameer2021,Zahedy2021,Lofthouse2023}, albeit these estimates are subject to the model assumptions.

The major limitation of absorption line studies is that they probe the gas only along a pencil-beam sightline towards the background point source and hence cannot directly map the spatial distribution of the gas. In contrast, spatially resolved emission can directly map the extent and morphology of the gas around galaxies. However, due to its low surface brightness, the CGM has been challenging to detect in emission at cosmological distances. Now, thanks to sensitive integral field unit (IFU) spectrographs, such as the Multi Unit Spectroscopic Explorer \citep[MUSE;][]{Bacon2010} on the Very Large Telescope (VLT), it is becoming possible to detect extended \lya\ emission \citep[e.g.][]{Borisova2016,Leclercq2017,Arrigoni2019,Fossati2021}, and metal line emission \citep[e.g.][]{Burchett2021,Zabl2021,Leclercq2022,Dutta2023} around galaxies, although individual detections of metal line emission from the CGM are still only a handful and limited to the inner, most dense, CGM. 

Alternative approaches to study the spatial extent and coherence scale of the CGM have been to use quasar pairs, gravitationally lensed quasars and arcs, and extended background galaxies. Absorption lines detected towards close quasar pairs have been used to measure the physical extent of the metal-enriched CGM using transverse clustering analysis over hundreds of kpc \citep[e.g.][]{Hennawi2007,Martin2010,Rubin2015,Mintz2022}. Multiple images of gravitationally lensed quasars have been used to study the structure of the absorbing gas on smaller length scales of few kpc \citep[e.g.][]{Rauch2001,Chen2014,Zahedy2016,Rubin2018,Kulkarni2019,Augustin2021}. Estimates of the coherence length scale, i.e. the length scale over which absorption does not vary significantly, from these studies suggest that the high-ionization gas phase is more coherent and spread out than the low-ionization gas phase. Recently, it has become possible to probe the absorbing gas over an extended region at sub-kpc scales using MUSE observations of background lensed arc images and extended galaxies \citep[e.g.][]{Lopez2018,Peroux2018,Tejos2021}. These studies have found that the distribution of metals around individual galaxies is similar to that around statistical samples of galaxies, and that there are indications of metals being mixed efficiently on kpc-scales in the CGM.

However, systems in which tomographic studies can be conducted are rare, and the structure of the CGM at different length scales is still poorly constrained. It is, therefore, important to expand the sample of systems in which one can spatially map the metal-enriched gas. In this work, we study the metal absorption towards multiple images of two background quasars that are gravitationally lensed by galaxy clusters. Thanks to strong gravitational lensing (magnification $\approx$2.5--17), the MUSE spectra of the multiple quasar images are among the ones with the largest equivalent exposure times ($\approx$30--1400 h) available to date. Unlike most lensed quasar studies that typically probe smaller physical separations ($\lesssim$20\,kpc), the angular separations ($\approx$2--22 arcsec) between the different pairs of quasar images studied here enable us to probe the absorbing gas over a wide range of physical separations, $\approx$0.4-150\,kpc. The MUSE spectroscopic observations of these two fields enable constructing an accurate lens mass model and, consequently, accurate estimates of the physical separation between quasar sightlines at the absorber redshifts. We use the \mgii\ and \feii\ absorption lines to study the coherence of the low-ionization gas phase and the \civ\ absorption lines to study the coherence of the high-ionization gas phase. Furthermore, the MUSE observations of the two lensed systems allow us to study the association of metal absorption with nearby galaxies.

The paper is structured as follows. We provide a brief overview of the observations used in this work in Section~\ref{sec_analysis}. The results on the correlation of metal absorption lines between different quasar sightlines and on the connection with galaxies are presented in Section~\ref{sec_result}. We discuss and summarise the results in Section~\ref{sec_discussion}. We adopt a Flat $\Lambda$CDM cosmology with $H_0 = 70$\,\kms\ Mpc$^{-1}$ and $\Omega_M = 0.3$, and express all distances in physical units.

\begin{figure*}
    \centering
    \includegraphics[width=0.44\textwidth, trim={1cm 0cm 3cm 1cm}, clip=True]{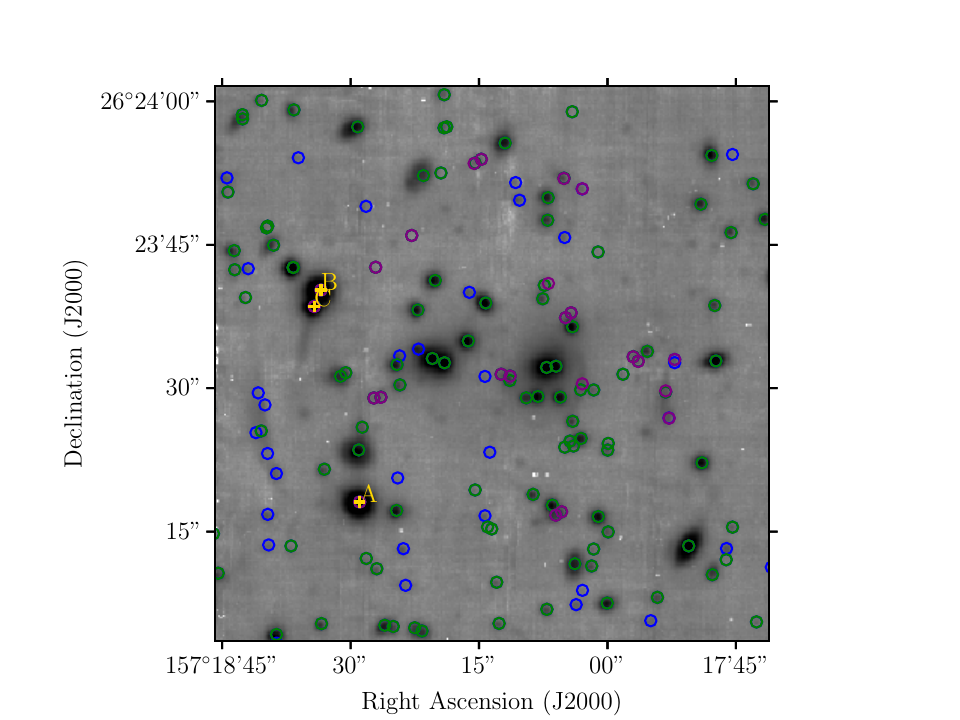}
    \includegraphics[width=0.55\textwidth]{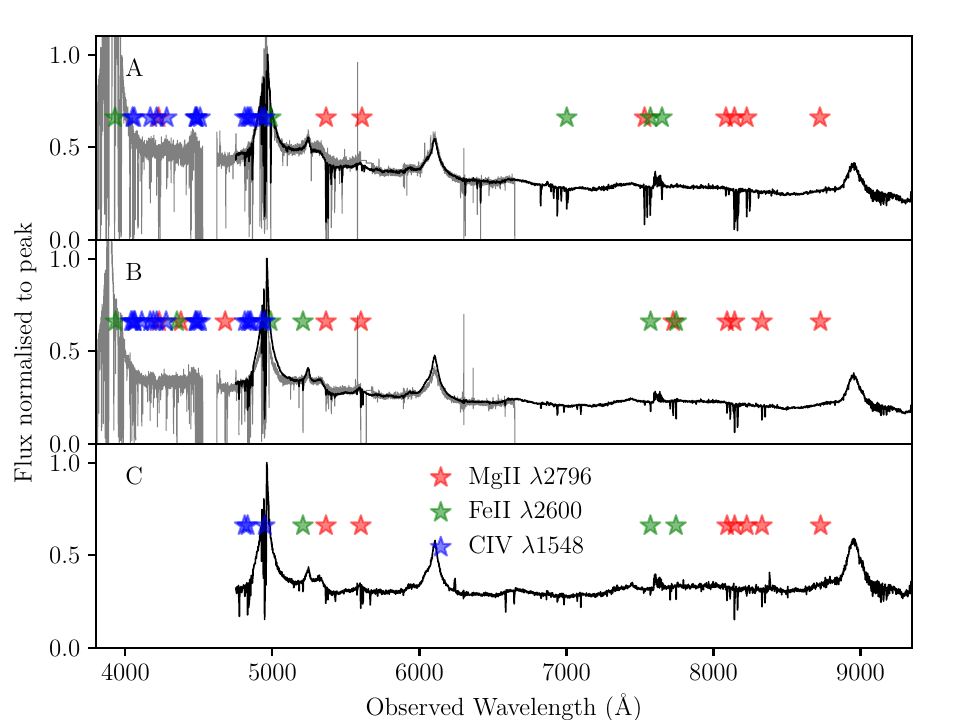}
    \caption{{\it Left:} MUSE white light image of the field J1029. The three multiple images, A, B, and C, of the quasar are marked with `+'. The blue circles identify all the sources detected in the HST F814W image, the green circles denote the sources that have reliable spectroscopic redshifts, and the purple circles indicate the multiple imaged sources.  
    {\it Right:} Spectra of the three quasar multiple images extracted from the MUSE cube. The UVES spectra are also shown for images A and B in grey. The stars mark the \mgii\ $\lambda$2796 (red), \feii\ $\lambda$2600 (green), and \civ\ $\lambda$1548 (blue) absorption lines detected in the spectra.}
    \label{fig:j1029}
\end{figure*}

\begin{figure*}
    \centering
    \includegraphics[width=0.44\textwidth, trim={1cm 0cm 3cm 1cm}, clip=True]{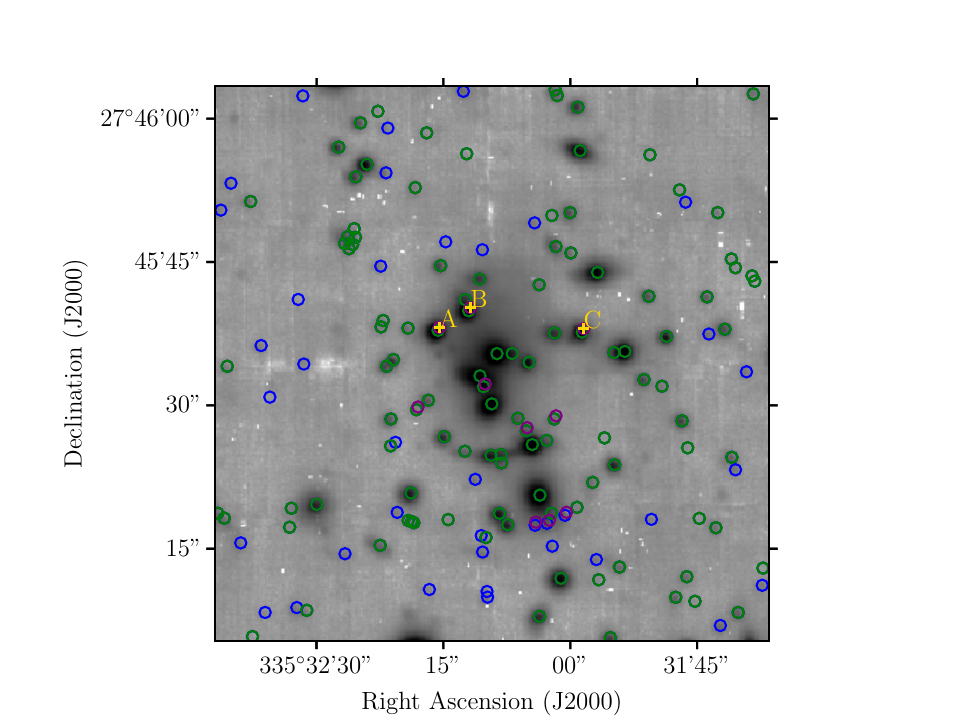}
    \includegraphics[width=0.55\textwidth]{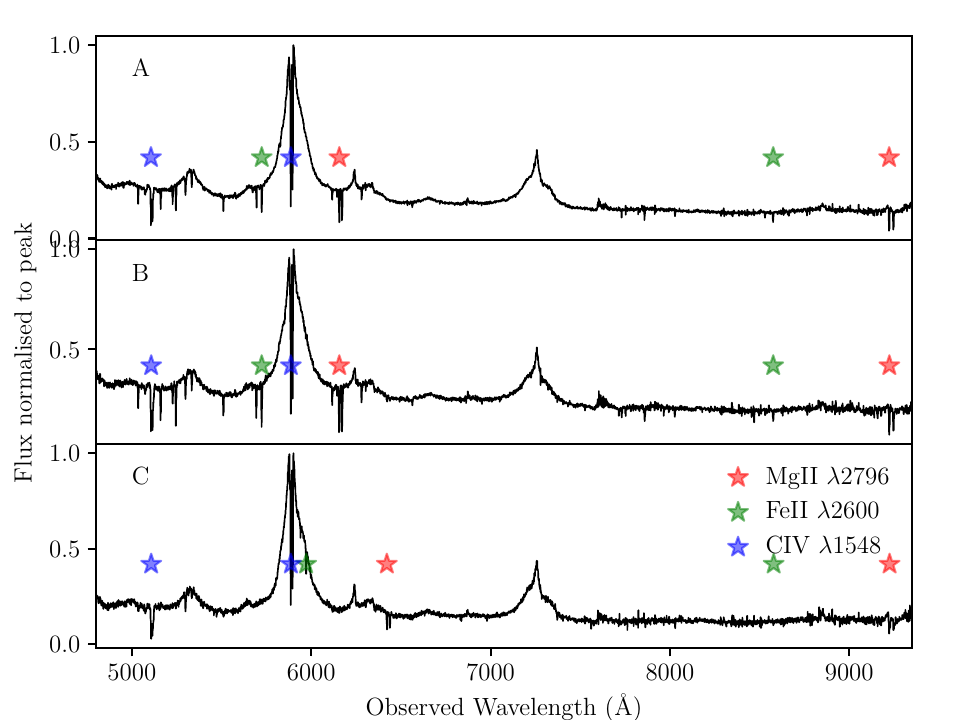}
    \caption{Same as in Fig.~\ref{fig:j1029} for the field J2222.}
    \label{fig:j2222}
\end{figure*}

\section{Observations \& Analysis}
\label{sec_analysis}

The fields SDSS J1029$+$2623 (J1029 hereafter) and SDSS J2222$+$2745 (J2222 hereafter) were observed with MUSE for 4.8 h and 4.4 h on target, respectively [PID: 0102.A-0642(A), 0103.A-0554(A); P.I.: C. Grillo]. The observations took place between 2019 March and July. Based on the analysis of the MUSE Analysis of Gas around Galaxies survey \citep{Lofthouse2023}, which consists of similar exposure times, we estimate the MUSE data to have a 90 percent photometric completeness limit of $r$ $\approx26$ mag for point sources, and a 90 percent spectroscopic flux completeness limit of $\approx 3 \times 10^{-18}$\,\ergscm\ at 5000 \AA\ for point sources. The data were reduced using the ESO pipeline \citep{Weilbacher2020} following standard procedures. The sky subtraction was performed using the Zurich Atmosphere Purge \citep{Soto2016}. Further details of the data reduction are provided in \citet{Acebron2022a} and \citet{Acebron2022b}. The wavelength coverage of the MUSE data is $\approx4750-9350$\,\AA, and the spectral resolution is $\approx2000-3000$.  

The sources are detected based on the Hubble Space Telescope (HST) WFC3/F814W image using {\sc SExtractor} \citep{Bertin1996}. The HST observations and data reduction are described in \citet{Oguri2013} and \cite{Sharon2017}. The one-dimensional spectra of the sources detected in HST were extracted from the MUSE cubes within a circular aperture of radius 0.8 arcsec, similar to the full-width at half-maximum of the point spread function. The redshifts of the sources were estimated by cross-matching the spectra with different spectral templates and emission lines. In this work, we consider the galaxies with reliable redshifts \citep[quality flag $\ge2$;][]{Acebron2022a,Acebron2022b}. 

The left panel of Fig.~\ref{fig:j1029} shows the MUSE image of the field J1029. This field contains three multiple images (A, B, C) of a quasar at \zqso\ = 2.1992 that is lensed by a galaxy cluster at \zlen\ = 0.588. The angular separation between the quasar images ranges from 1.9 to 22.5 arcsec. The MUSE spectra of the three quasar multiple images are shown in the right panel of Fig.~\ref{fig:j1029}.

In addition to the MUSE spectra, archival [PID: 092.B-0512(A)] UVES spectra are available for the multiple images A and B of the quasar J1029. The UVES observations were performed from 2014 January to February. The reduced UVES spectra were obtained from the ESO archive. The individual exposure spectra were combined by interpolating to a common wavelength array and weighting each flux pixel by the inverse variance. The wavelength coverage of the final spectrum is $\approx3300-6600$\,\AA, and the spectral resolution is $\approx30000$. The UVES spectra of the quasar multiple images A and B, along with the MUSE spectra, are shown in the right panel of Fig.~\ref{fig:j1029}.

The field J2222 (see left panel of Fig.~\ref{fig:j2222}) contains six multiple images (A, B, C, D, E, F) of a background quasar at \zqso\ = 2.801 that is lensed by a galaxy cluster at \zlen\ = 0.489. The quasar images D, E and F are contaminated by the light from nearby bright cluster members in the MUSE data, and their one-dimensional spectra cannot be utilized for an absorption line analysis. This work exploits the spectra of the quasar images A, B, and C (see right panel of Fig.~\ref{fig:j2222}). The angular separation between these images ranges from 4 to 15 arcsec.

All the MUSE and UVES quasar spectra were converted into the heliocentric frame and vacuum wavelength for the analysis. For the identification and analysis of absorption lines, the continuum normalisation of the quasar spectra was carried out by spline fitting using {\sc linetools}. The signal-to-noise ratio per pixel of the MUSE quasar spectra ranges between $\approx$30 and $\approx$90 around 7000\,\AA, and that of the UVES quasar spectra is $\approx$20 around 5000\,\AA.

\section{Results}
\label{sec_result}

\begin{figure*}
    \centering
    \includegraphics[width=0.24\textwidth]{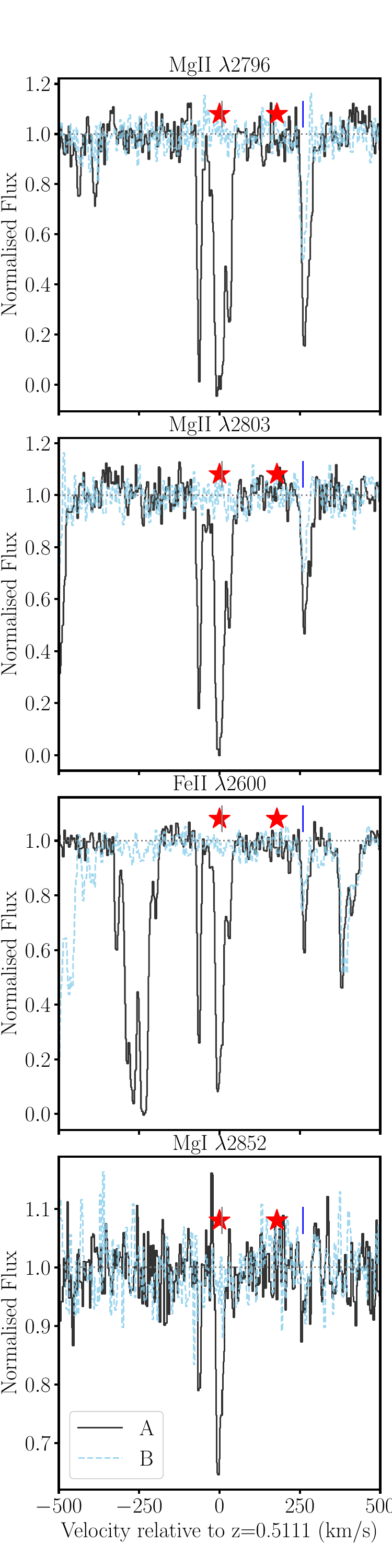}
    \includegraphics[width=0.24\textwidth]{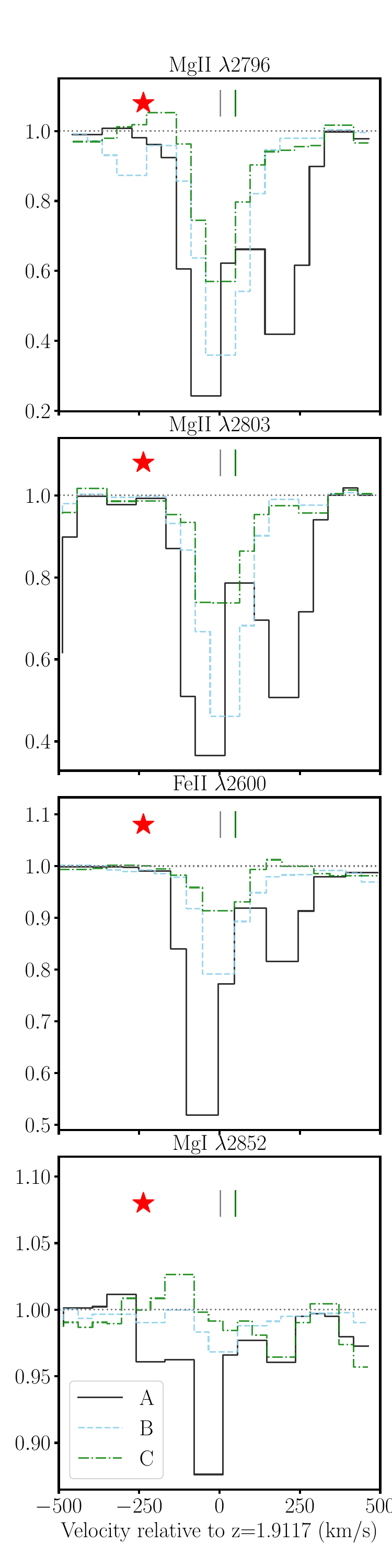}
    \includegraphics[width=0.24\textwidth]{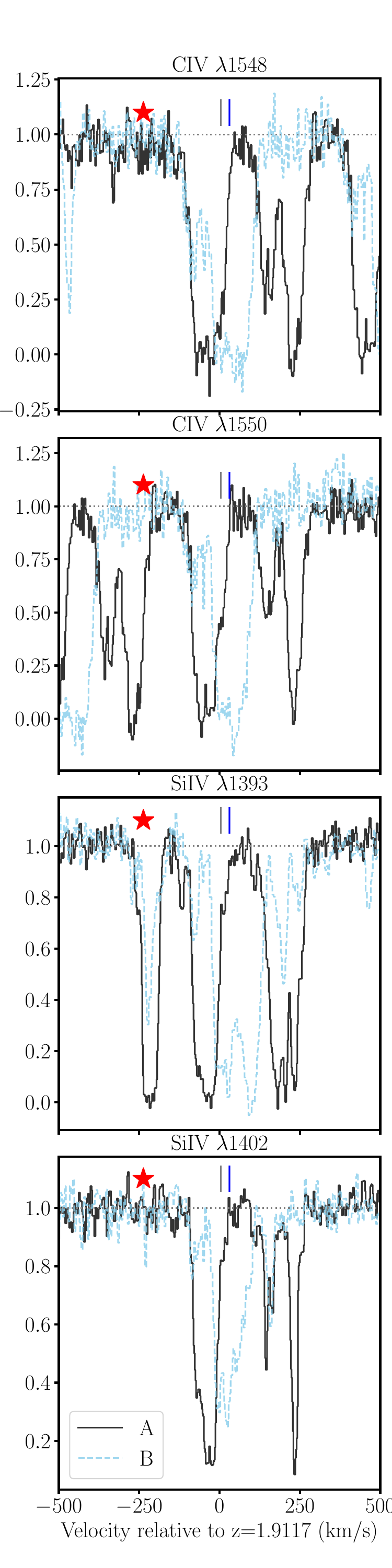}
    \includegraphics[width=0.24\textwidth]{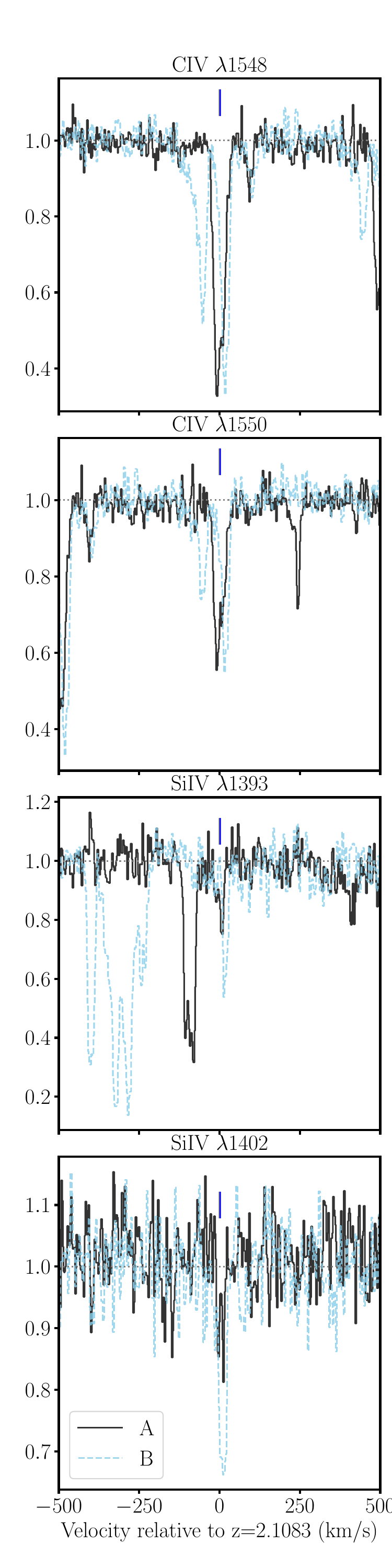}
    \caption{Examples of absorption lines detected towards the different multiple images, A, B, and C, of the quasar J1029. The corresponding ion is indicated at the top of each panel. The second column shows the normalised MUSE spectra of the multiple images A (black solid lines), B (blue dashed lines), and C (green dot-dashed dashed lines). The other columns show the normalised UVES spectra of the quasar images A (black solid lines) and B (blue dashed lines), respectively. The optical depth-weighted median redshifts of the absorbers are marked by vertical ticks (grey for A, blue for B, green for C). The relative velocities of the galaxies identified in MUSE are marked by red stars.}
    \label{fig:abs_lines}
\end{figure*}

\subsection{Metal absorption across two quasar sightlines}
\label{sec_results_abspair}

We identified all the metal absorption lines, such as \mgii, \mgi, \feii, \civ, and \sifour, in each of the quasar normalised spectra redward of the quasar \lya\ emission line. The \mgii, \feii, and \civ\ systems identified in the different quasar image spectra are listed in the appendix and also marked on top of the quasar spectra shown in the right panels of Figures~\ref{fig:j1029} and \ref{fig:j2222}. We estimate the optical depth-weighted median redshift and rest-frame equivalent width of the absorption lines. We note that in the case of J1029, the UVES spectra of the quasar multiple images A and B provide coverage of the \mgii\ and \feii\ lines at the redshift of the lens cluster, $z\approx0.588$. However, we do not detect any absorption from the cool low-ionization gas (rest equivalent width of \mgii\ $\le$0.05\,\AA) at the redshift of the cluster. For both quasars, \civ\ absorption is detected at the quasar redshift towards all the multiple images. However, we do not include the intrinsic absorption in the following analysis.

For each quasar, we search for counterpart absorbers within $\pm500$\,\kms\ towards the different images. For context, the escape velocity at the virial radius for a galaxy group at $z=1$ with halo mass, \mhalo\ $\approx10^{13}$\,\msun, is $\approx500$\,\kms. In most cases, the absorber pairs lie within $\pm100$\,\kms\ of each other. Some examples of absorption lines detected across the multiple quasar sightlines are shown in Fig.~\ref{fig:abs_lines}. We find 20 \mgii\ absorber pairs and 10 \feii\ absorber pairs over $z\approx0.5-2.3$, and 18 \civ\ absorber pairs over $z\approx1.6-2.3$. For similar redshift, line-of-sight velocity window, and equivalent width limits (0.01\,\AA\ for \mgii\ and 0.05\,\AA\ for \civ), and based on the observed number density of \mgii\ and \civ\ absorbers \citep{Mathes2017,Hasan2020}, we expect to detect at random $\approx$0.6 \mgii\ and $\approx$1.4 \civ\ absorber pairs, respectively, across two quasar sightlines.

To quantify the variation in absorption across two different sightlines ($X$ and $Y$), we estimate the fractional difference in the rest-frame equivalent width for each absorber pair as, 
\begin{equation}
    \Delta W_r = (W_r^X - W_r^Y)/W_r^X,
\end{equation}
where $W_r^X > W_r^Y$. In case of non-detection in one of the sightlines, we estimate the $3\sigma$ upper limit on the equivalent width at the same redshift as the absorber along the other sightline, for a velocity width of 100\,\kms\ \citep[typical of \mgii\ absorbers at these redshifts; e.g.][]{Dutta2020} using the signal-to-noise ratio. 

\begin{figure}
    \centering
    \includegraphics[height=0.3\textheight]{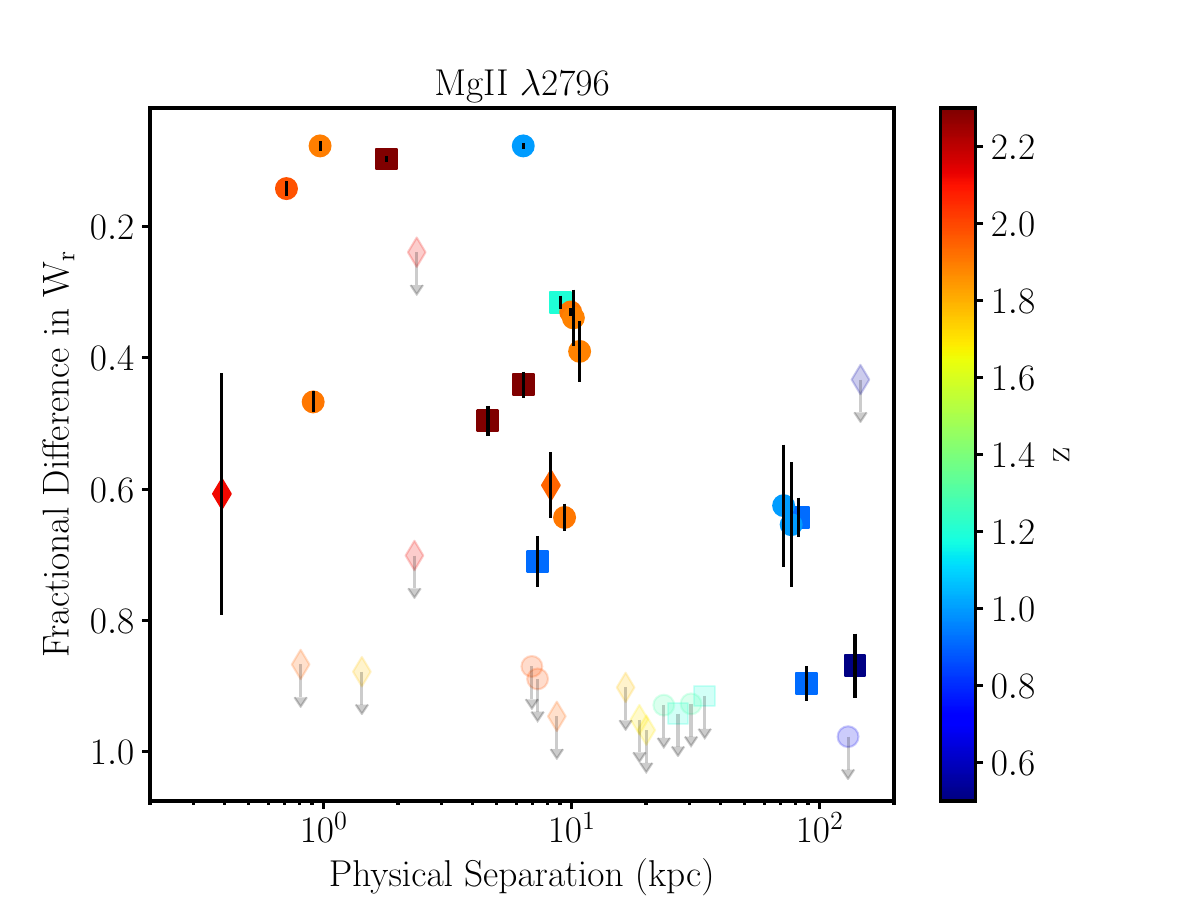}
    \includegraphics[height=0.3\textheight]{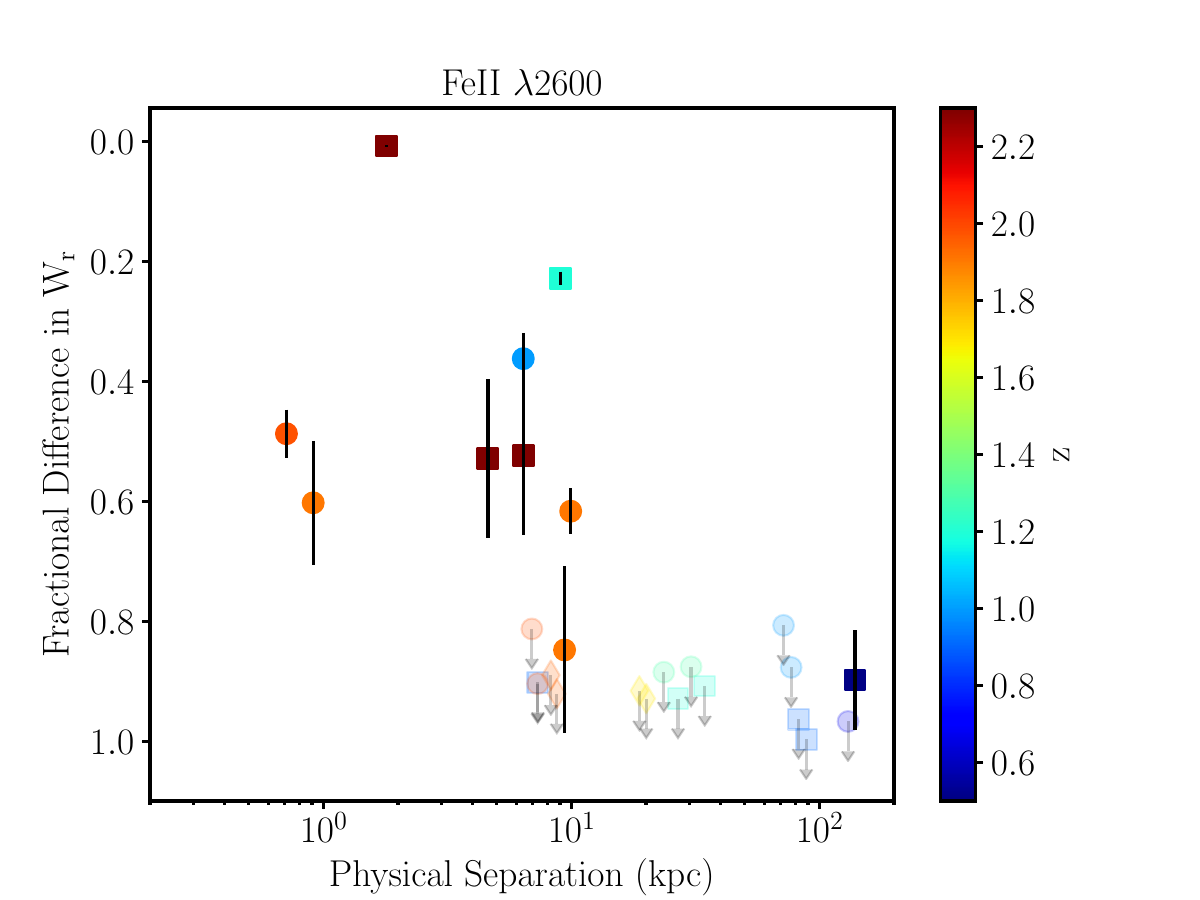}
    \includegraphics[height=0.3\textheight]{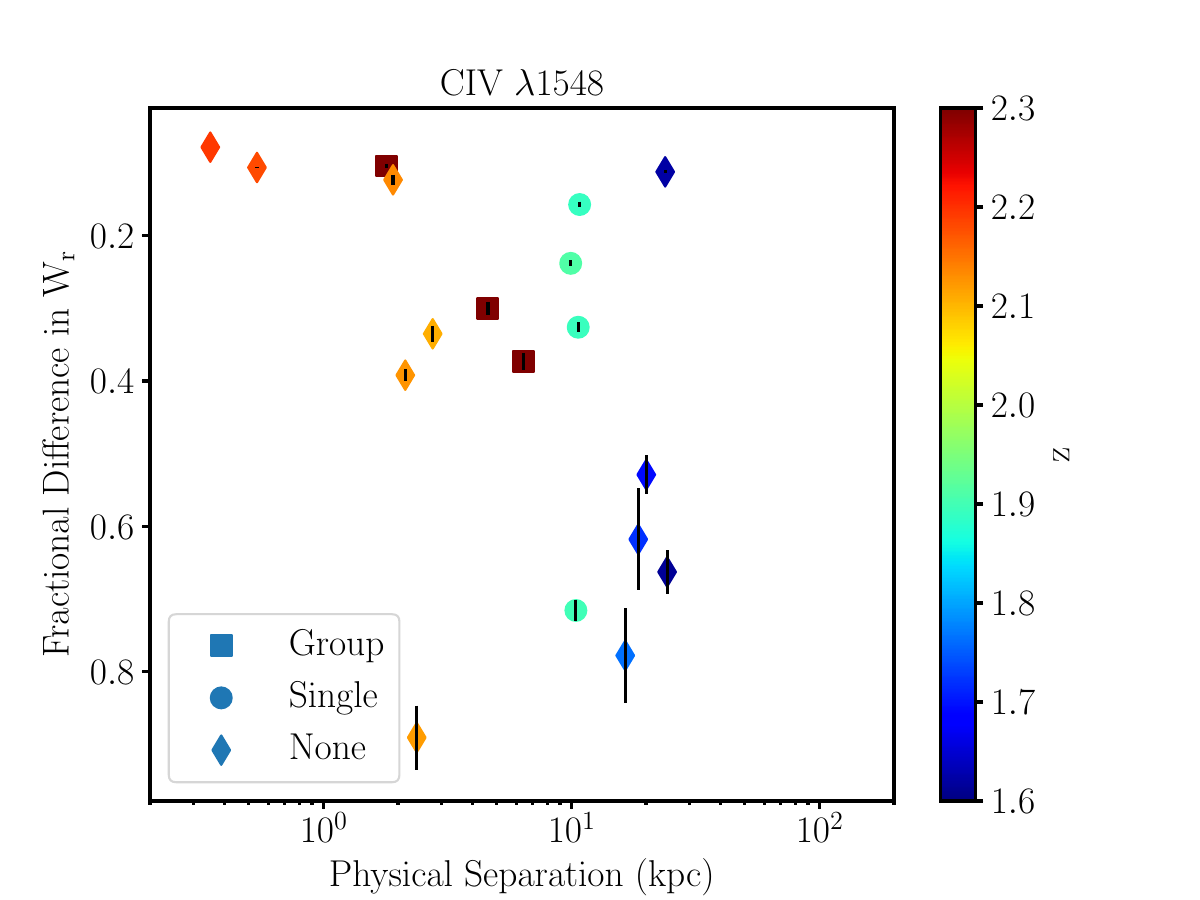}
    \caption{The fractional difference in equivalent width of \mgii\ (top), \feii\ (centre), and \civ\ (bottom) absorbers as a function of physical separation between two quasar multiple images. The markers are colored by the redshift of the absorbers. Square, circle, and diamond markers indicate groups, single, and no galaxies within $\pm500$\,\kms\ of the absorber redshift in the MUSE field-of-view, respectively.}
    \label{fig:diff_sep}
\end{figure}

To compute the transverse physical separation between two sightlines, we first ray-trace the positions of the quasar multiple images to the absorber redshifts using the best-fit lens models presented in \citet[][Model 1]{Acebron2022a} and in \citet[][ES-Model]{Acebron2022b}. For further details on these lens models, we refer to the above papers. Using the best-fit predicted positions of the quasar multiple images at different redshifts, we estimate the physical separation between the sightlines in the absorber plane. The uncertainties in the physical separation, at the median redshift of $\approx$1.8 of the sample, due to uncertainties in the lens modeling, range from $\approx$0.4 kpc to $\approx$1 kpc with an average of $\approx$0.8 kpc for J1029, and from $\approx$0.8 kpc to $\approx$4 kpc with an average of $\approx$3 kpc for J2222. The fractional difference in equivalent width as a function of the physical separation for the \mgii\ $\lambda$2796, \feii\ $\lambda$2600, and \civ\ $\lambda$1548 intervening absorption lines are shown in Fig.~\ref{fig:diff_sep}. 

In the case of the \mgii\ and \feii\ absorber pairs, we are able to probe over a large range of transverse physical separations, $\approx$0.4--150\,kpc, while we probe up to $\approx$25\,kpc separations for the \civ\ absorber pairs. The fractional difference in equivalent width, $\Delta W_r$, shows a positive correlation with the physical separation. Considering the detections, for \mgii, the Spearman rank order correlation coefficient is $r_s$ = 0.58, and the probability of the correlation arising by chance or $p$-value = 0.007, for \feii, $r_s$ = 0.55, $p$-value = 0.09, and for \civ, $r_s$ = 0.49, $p$-value = 0.04, with similar results obtained considering the limits as detections. This increasing trend of $\Delta W_r$ with physical separation indicates that, as expected, the metal absorption becomes less spatially coherent over larger separations. We highlight specifically a transition around physical separation of $\approx$10 kpc, particularly for \mgii\ and \feii, beyond which the fractional difference shows a considerable drop. Based on this we suggest a coherence length scale of $\approx$10 kpc for the metal-enriched gas around galaxies.

However, as can also be seen from Fig.~\ref{fig:diff_sep}, there is a considerable scatter in $\Delta W_r$ ($\Delta W_r$ ranges from $\approx$0.08 to $\approx$0.9) over the full range of physical separations, suggesting large variations in the spatial coherence. Particularly for \mgii, we see non-detections of paired absorption over almost the full range of physical separation probed. On the contrary, \civ\ absorption is always detected in both sightlines, albeit we probe up to only $\approx$25\,kpc in this case. On average, the high-ionization gas phase shows a slightly higher level of coherence than the low-ionization gas phase, with \civ\ systems showing average $\Delta W_r$ $\approx$0.3 compared to average $\Delta W_r$ $\approx$0.4 for \mgii\ systems, and $\Delta W_r$ $\approx$0.5 for \feii\ systems, at a separation of less than 25 kpc.

\begin{figure}
    \centering
    \includegraphics[height=0.3\textheight]{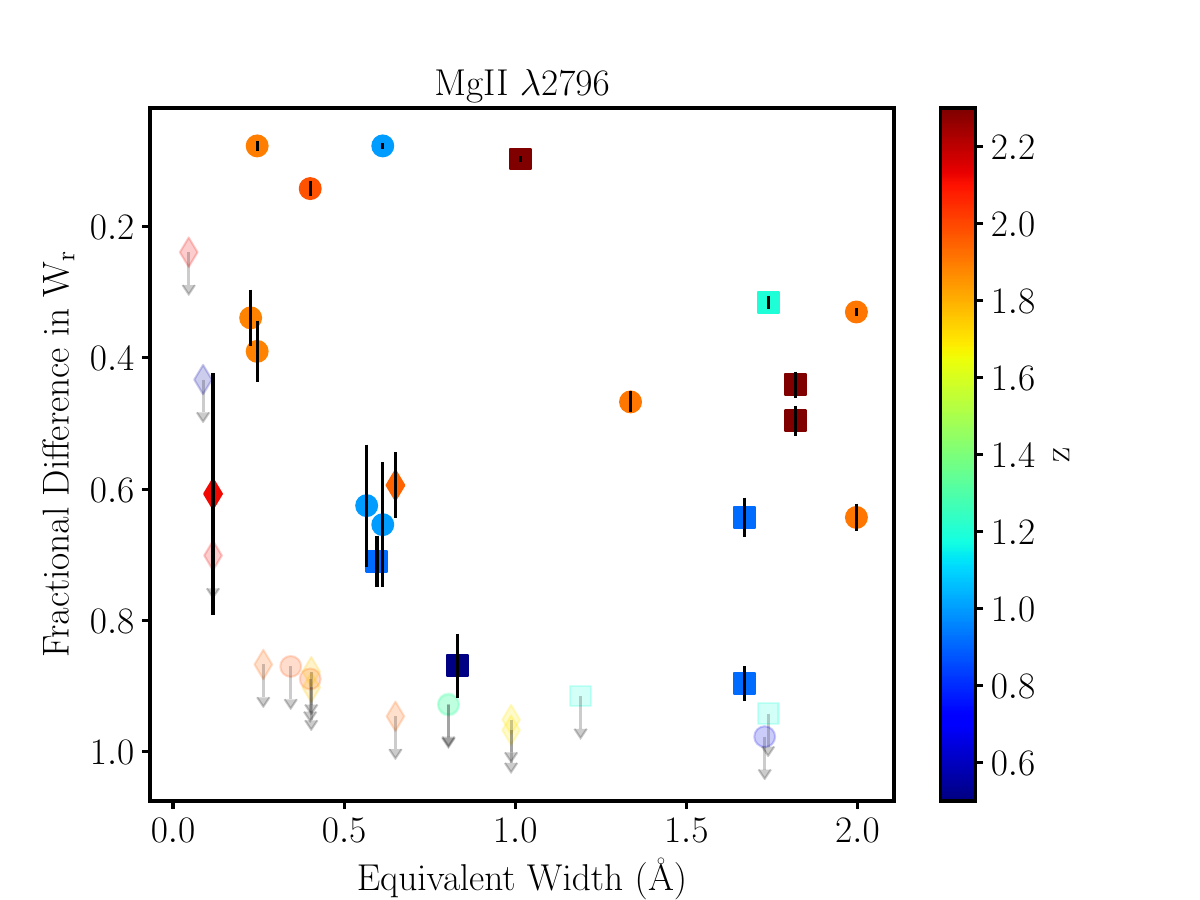}
    \includegraphics[height=0.3\textheight]{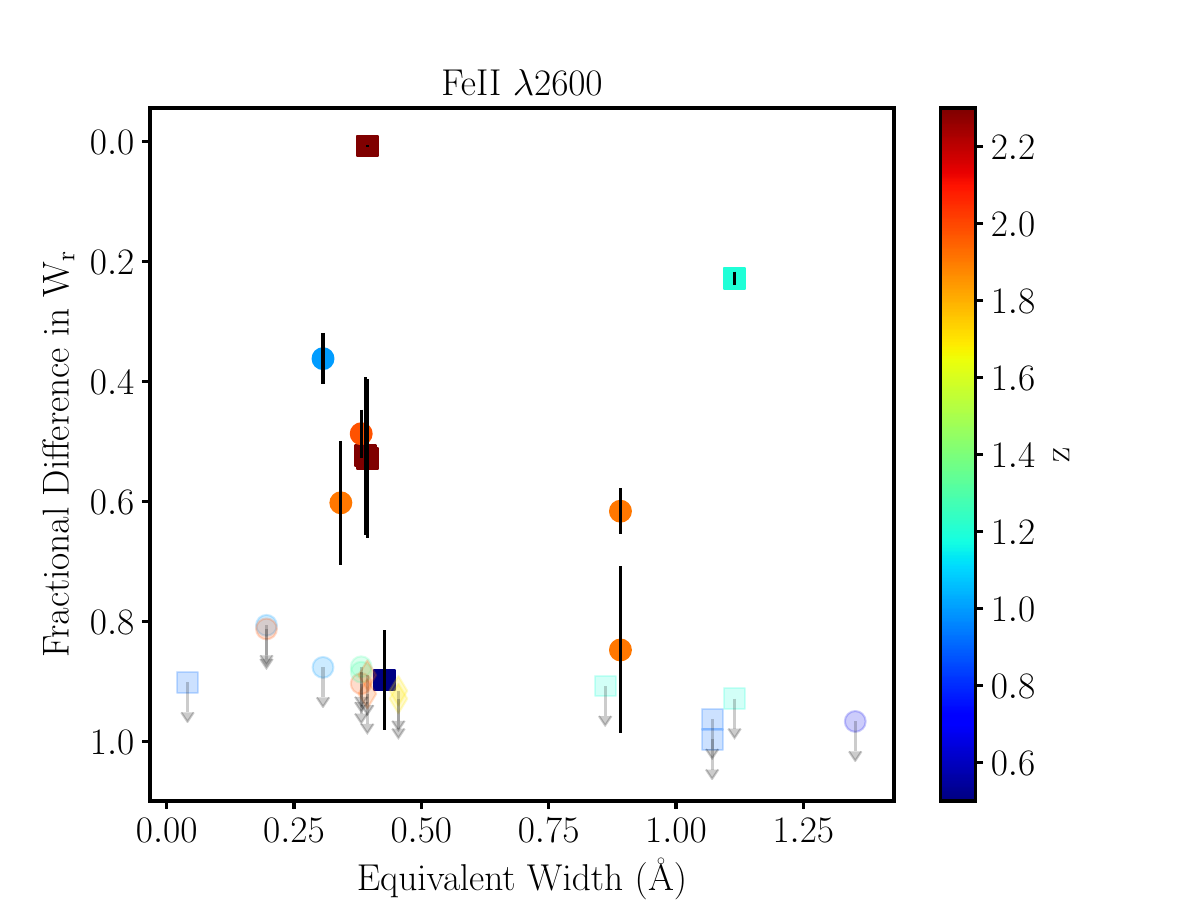}
    \includegraphics[height=0.3\textheight]{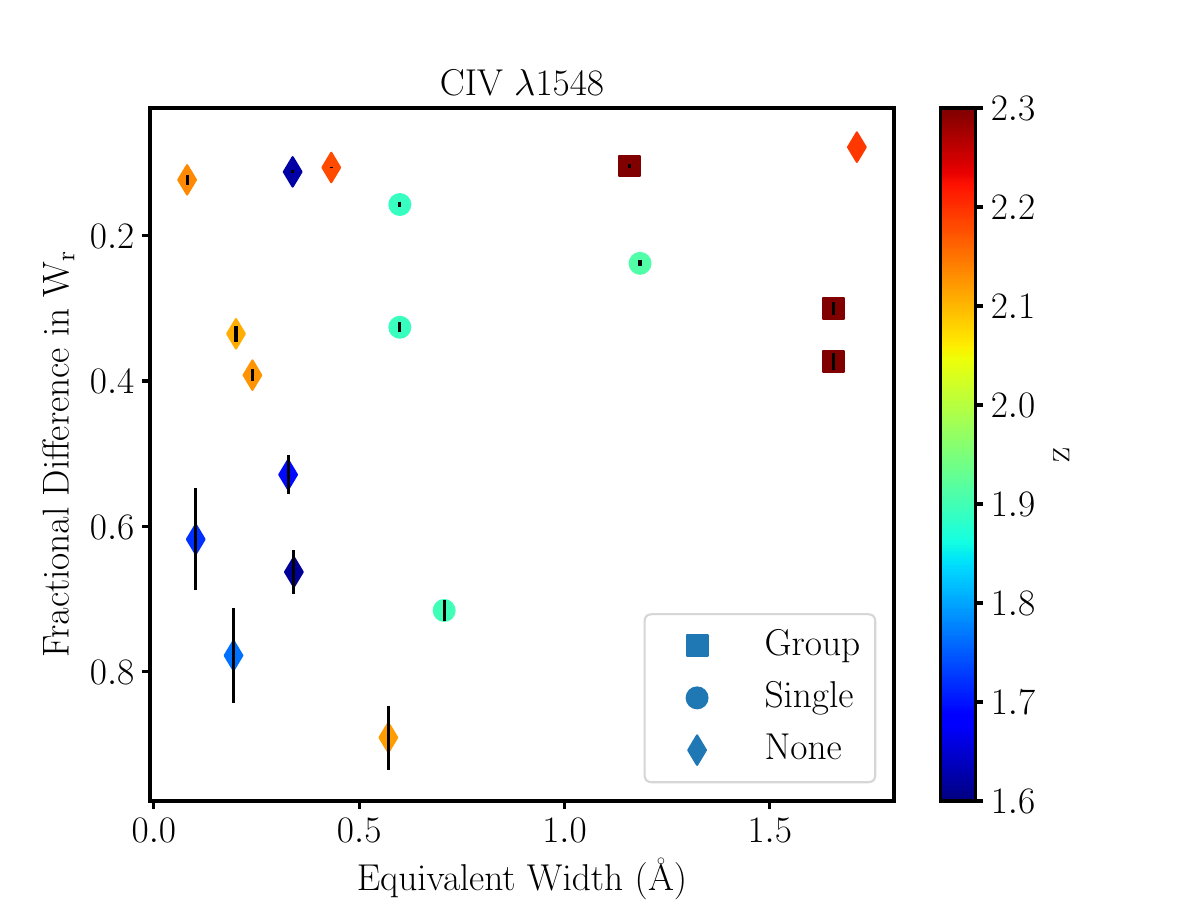}
    \caption{The fractional difference in equivalent width of \mgii\ (top), \feii\ (centre), and \civ\ (bottom) absorbers as a function of the equivalent width of the stronger absorber in a pair. The symbols are the same as in Fig.~\ref{fig:diff_sep}.}
    \label{fig:diff_ew}
\end{figure}

Fig.~\ref{fig:diff_ew} shows the fractional difference in the equivalent width of the metal absorbers as a function of the equivalent width of the stronger absorber in a pair. We do not find any significant dependence of $\Delta W_r$ on the equivalent width for \mgii, \feii, and \civ\ absorbers, suggesting that coherence is independent of the absorption strength. Additionally, we do not find any significant dependence of $\Delta W_r$ on the absorber redshift. This indicates that the correlation found between metal absorption in this sample is primarily driven by the physical separation between the sightlines.

\begin{figure}
    \centering
    \includegraphics[width=0.5\textwidth]{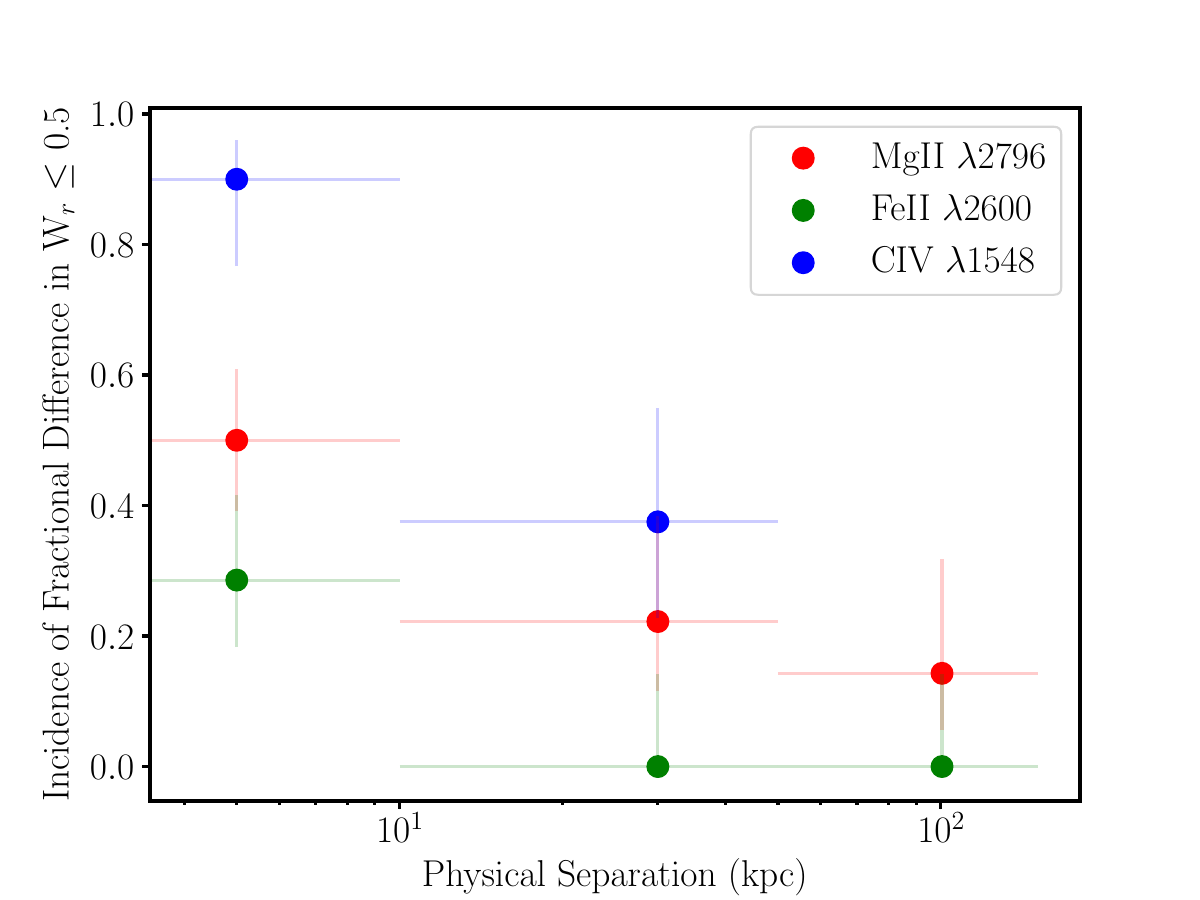}
    \caption{Incidence of fractional difference in equivalent width less than 0.5 as a function of the physical separation. Red, green, and blue markers show the incidence for \mgii, \feii, and \civ\ absorbers, respectively. The vertical bars show 1$\sigma$ Wilson score confidence intervals on the incidence. The range in physical separation included in each bin is shown by the horizontal bars.}
    \label{fig:diff_frac}
\end{figure}

To further investigate the coherence of the metal absorption as a function of physical separation, we plot in Fig.~\ref{fig:diff_frac} the incidence of $\Delta W_r$ $\le$0.5 in three different physical separation bins, $0-10$ kpc, $10-50$ kpc, and $50-150$ kpc. We consider the 3$\sigma$ limits as detections for this analysis, however, the relative trends do not change if we consider only the detections. The trends also remain similar when considering variations on the separation bins and coherence levels. 

Consistent with the results from Fig.~\ref{fig:diff_sep}, the incidence decreases with increasing physical separation for the \mgii, \feii, and \civ\ absorbers. The drop in incidence by a factor of $\approx$2 at separations larger than 10 kpc further supports this length scale being a characteristic coherence scale. \civ\ absorbers show the highest incidence of coherence, with about 90 percent of absorbers having $\Delta W_r$ $\le$0.5 at separations $\le$10 kpc, and $\approx$40 percent having $\Delta W_r$ $\le$0.5 over 10--50 kpc separation. This sample lacks sightline pairs that probe \civ\ at larger separations. \mgii\ absorbers exhibit a factor of two lower incidences than \civ, and a factor of two higher incidences than \feii\ at separations less than 10 kpc. About 50 percent of the \mgii\ absorbers have $\Delta W_r$ $\le$0.5 at separations $\le$10 kpc, and the fraction drops to $\approx$22 percent at 10--50 kpc, and $\approx$14 percent at 50--150 kpc. The presence of paired \mgii\ absorption, even at large separations, could represent coherent structures composed of cool gas clouds extending over an area of several tens of kpc$^2$. However, detailed comparison of the chemical composition and kinematic structure of the absorption is required to confirm this.of

\subsection{Metal absorption across three quasar sightlines}
\label{sec_results_abstriplet}

\begin{figure*}
    \centering
    \includegraphics[width=0.49\textwidth]{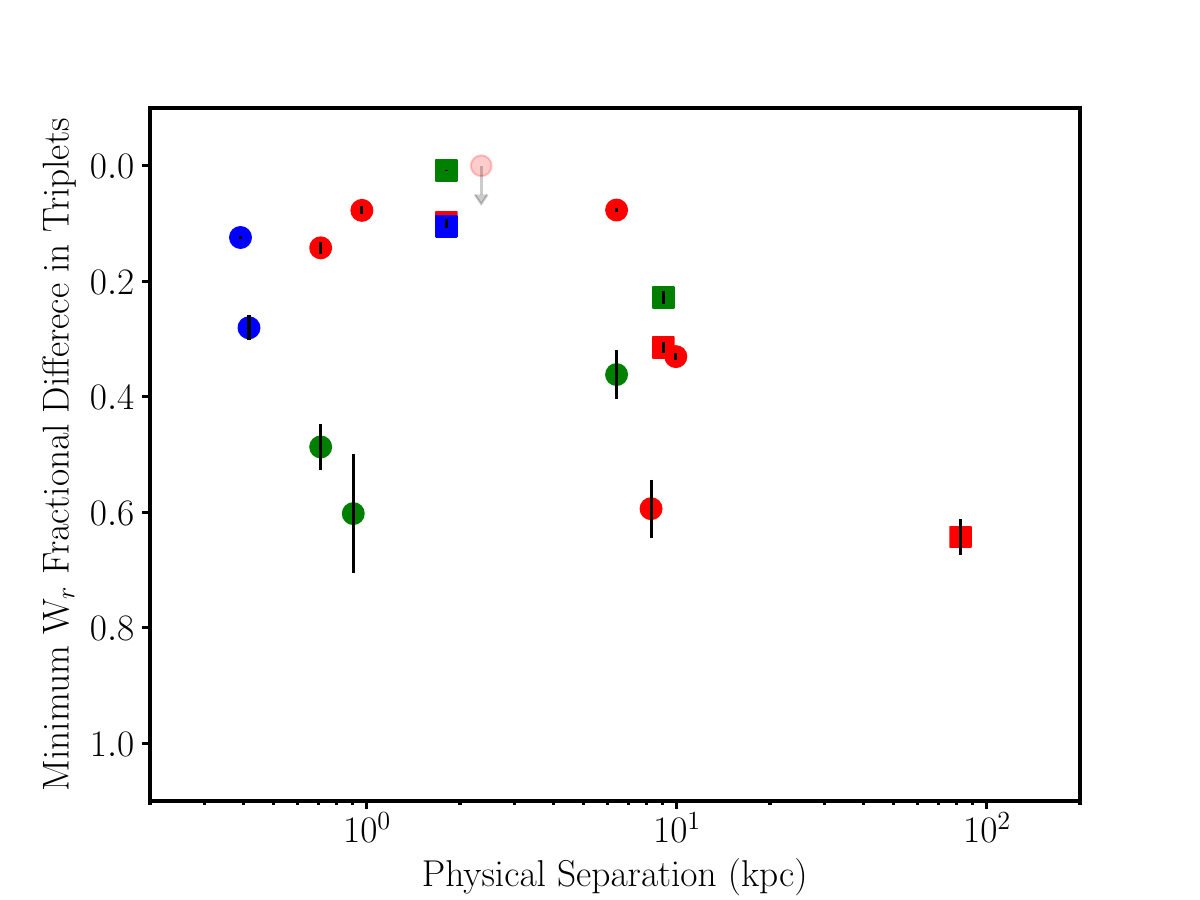}
    \includegraphics[width=0.49\textwidth]{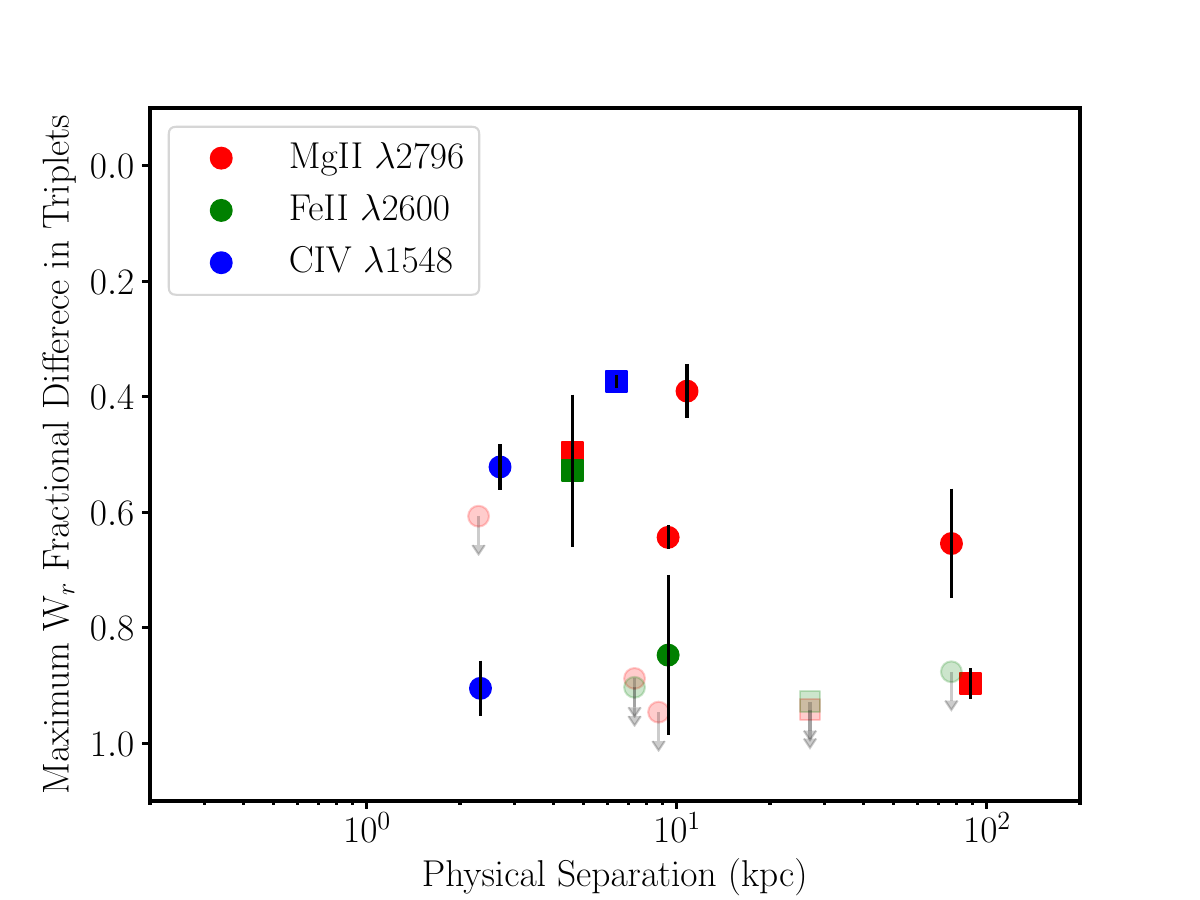}
    \caption{The minimum (left) and maximum (right) fractional difference in equivalent width of \mgii\ (red), \feii\ (green), and \civ\ (blue) absorbers between two quasar multiple images in a triplet system as a function of the physical separation between the images. Square and circle markers indicate that the absorption system is associated with group and single galaxies, respectively.}
    \label{fig:triple_abs}
\end{figure*}

Taking advantage of the presence of three quasar multiple images in each of the fields, we next look at the coherence of metal absorption across three different sightlines. Same as when searching for paired absorption in the previous section, we search for absorbers that are detected in all three quasar sightlines within $\pm$500\,\kms\ of each other along the line-of-sight direction. We find in total five \mgii\ absorber triplets, two \feii\ absorber triplets, and three \civ\ absorber triplets. For comparison, we expect to detect $\approx$0.005 and $\approx$0.03 \mgii\ and \civ\ absorber triplets, respectively, in three random quasar sightlines for the above velocity window and median redshift of the sample, based on the observed number density of \mgii\ and \civ\ absorbers \citep{Mathes2017,Hasan2020}. 

Similar to the analysis in Section~\ref{sec_results_abspair}, we quantify the variation in absorption across the three different sightlines ($X$, $Y$, and $Z$) by estimating the fractional difference in equivalent width or $\Delta W_r$ for each pair of sightlines, $XY$, $XZ$, and $YZ$. The minimum and maximum of the $\Delta W_r$ estimates among the three pairs are plotted as a function of the corresponding physical separation in Fig.~\ref{fig:triple_abs}. Given the limited sample size, we consider $\Delta W_r$ of all three ions together for the below discussion.

If absorption of similar strength was present across all three sightlines, then the minimum and maximum $\Delta W_r$ distributions would have been similar. Instead we find that the distributions are significantly different, with the minimum $\Delta W_r$ being $\approx$0.2 on average, and the maximum $\Delta W_r$ being $\approx$0.8 on average. A two-sample Kolmogorov–Smirnov (KS) test gives the maximum difference between the cumulative distributions of the two samples as $D$ = 0.8 with a $p$-value of $3\times10^{-5}$. Moreover, the minimum $\Delta W_r$ occurs between sightlines with smaller physical separations (median of $\approx$2 kpc) compared to the maximum $\Delta W_r$ (median of $\approx$9 kpc). The distributions of the two physical separations are also different based on a KS test ($D$ = 0.5, $p$-value = 0.01). 

The above tells us that typically metal absorption of similar strength are detected across two close ($\lesssim$10 kpc) quasar sightlines, and when detected towards a third sightline further apart, the metal absorption strength changes significantly. In other words, consistent with the results found in Section~\ref{sec_results_abspair}, the metal absorption becomes less coherent with increasing physical separation. Combining the minimum and maximum $\Delta W_r$ estimates, there is a clear correlation of $\Delta W_r$ with physical separation ($r_s$ = 0.57, $p$-value = $5\times10^{-4}$).

\subsection{Association of galaxies and metal absorption}
\label{sec_results_gal}

\begin{figure*}
    \centering
    \includegraphics[width=0.33\textwidth, trim=0 0 0.7cm 0, clip=True]{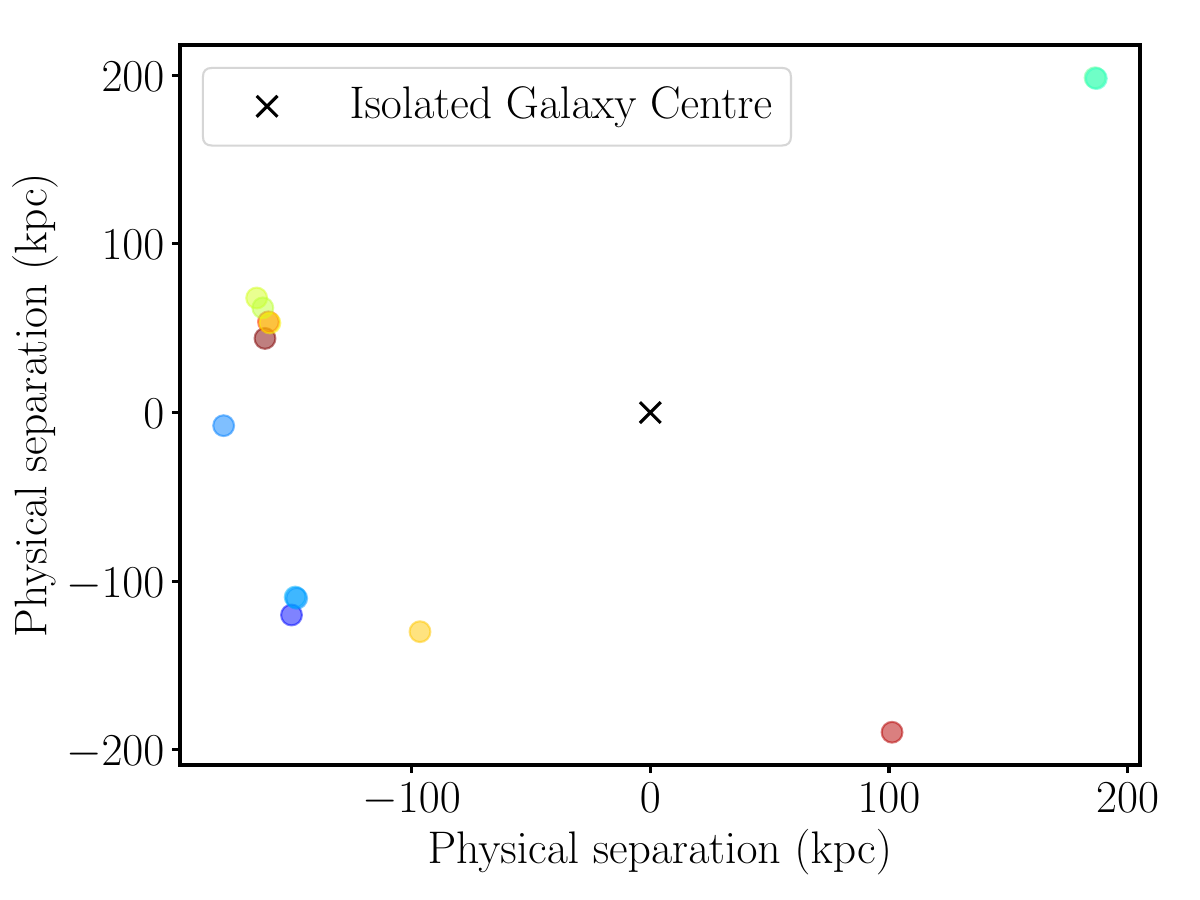}
    \includegraphics[width=0.31\textwidth, trim=1cm 0 1cm 0, clip=True]{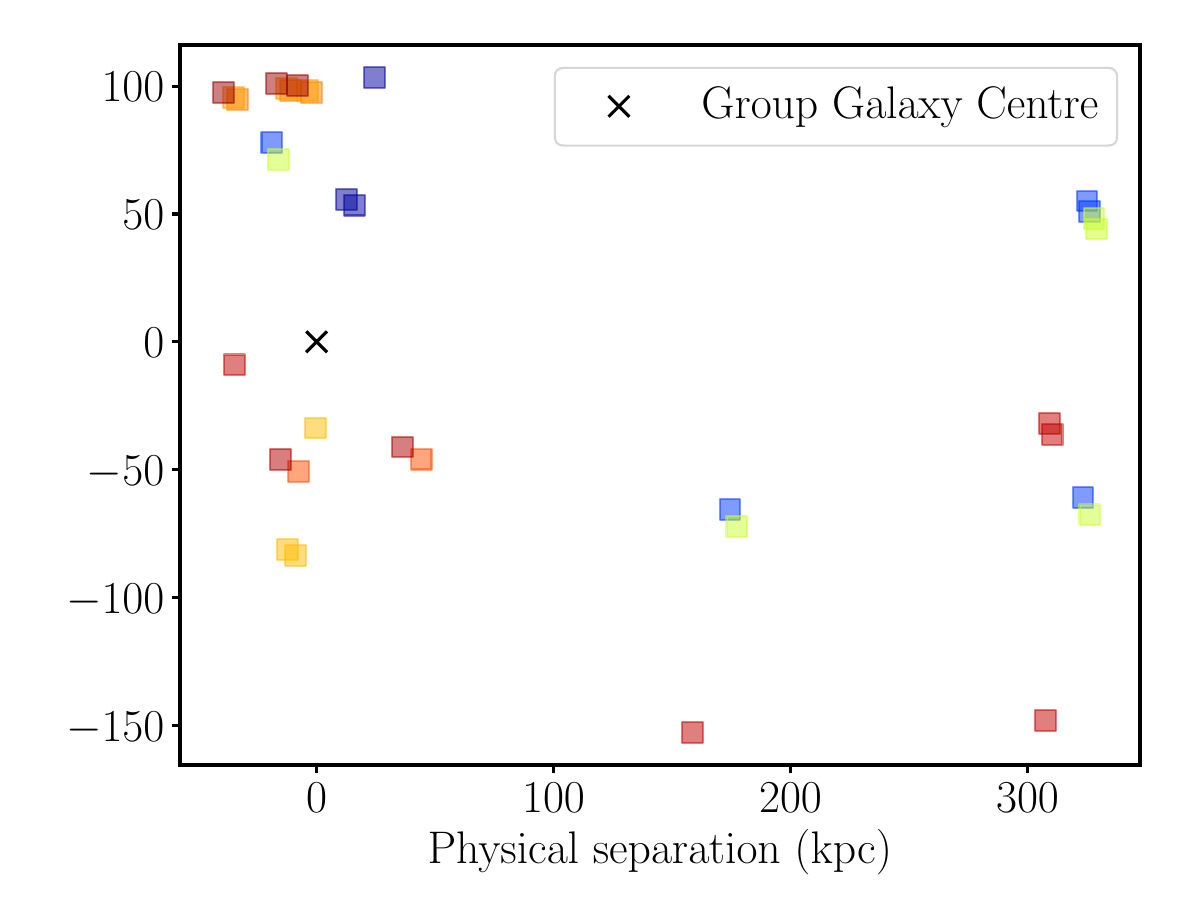}
    \includegraphics[width=0.33\textwidth, trim=1cm 0 0 0, clip=True]{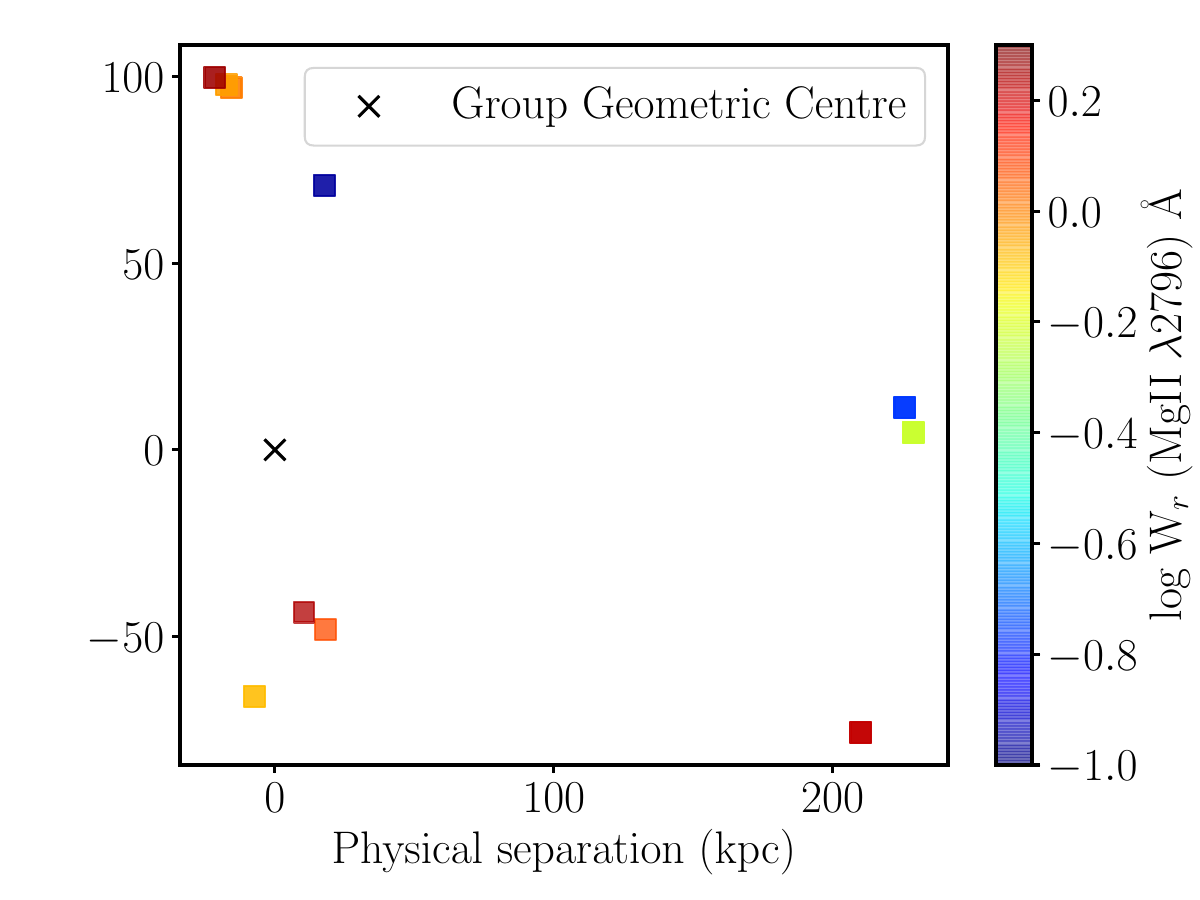}
    \caption{The locations of the \mgii\ absorbers around the isolated galaxies (left), group galaxies (centre), and geometric centre of the groups (right). The $x$ and $y$ axes are the physical separations in RA and Dec, respectively, with respect to the galaxy/group in the absorber plane. The markers are colored by the equivalent width of the \mgii\ absorbers.}
    \label{fig:gal_abs}
\end{figure*}

When studying the galaxy-diffuse gas connection, the advantage of MUSE IFU data is that they allow us to simultaneously study the galaxy environment around the gas detected in absorption towards the quasar multiple images. Having investigated the coherence of metal absorption lines across the multiple quasar images, we look next at how the absorption is connected to galaxies. To do this, we search for galaxies within $\pm500$\,\kms\ of the absorber redshift in the MUSE field-of-view ($\approx500\times500$ kpc$^2$ at $z\approx1$). 

In Figures \ref{fig:diff_sep} and \ref{fig:diff_ew}, the different markers indicate whether the absorber pair detected across two quasar sightlines is associated with a single, more than one, or no galaxies. For almost all the \mgii\ (90 percent) and \feii\ (100 percent) absorber pairs, there is at least one galaxy counterpart detected in MUSE. About 40 and 50 percent of the \mgii\ and \feii\ absorber pairs are associated with more than one galaxies (loosely termed as groups henceforth), respectively. For only about 40 percent of the \civ\ absorber pairs, we are able to identify galaxies. This is because at the redshifts of the \civ\ absorbers ($z\approx1.5-2$) that can be detected redward of the quasar \lya\ forest, it is typically challenging to identify galaxies due to the lack of strong emission lines in the optical wavelengths. When metal absorption is detected across all three quasar sightlines, we find that there is always at least one associated galaxy, with $\approx$60 percent of the systems being associated with a single galaxy, and rest with a group of galaxies (Fig.~\ref{fig:triple_abs}).

There appears to be no significant difference between groups and single galaxies regarding the distribution of the fractional difference in the equivalent width of the associated metal absorption, albeit the sample sizes are statistically small. For the \mgii\ absorber pairs, which have the largest sample size, we estimate the incidence of $\Delta W_r$ $\le$0.5 similar to Fig.~\ref{fig:diff_frac}, but for groups and single galaxies separately, in two physical separation bins, $0-10$ kpc and $10-150$ kpc. We do not find any significant difference in the incidence between \mgii\ absorbers associated with groups and single galaxies. However, given the small sample sizes, which hinder controlling for other parameters such as physical separation and redshift, it is difficult to derive any physical implication from the above.

Furthermore, we combined all the \mgii\ absorbers detected towards the different quasar multiple images and looked at how they are distributed around the associated galaxies (i.e. those within $\pm500$\,\kms\ and in the MUSE field-of-view). Fig.~\ref{fig:gal_abs} shows the locations of the \mgii\ absorbers around the isolated and group galaxies, and also around the geometric centre of the groups. We again do not find any significant difference in the distribution of the \mgii\ absorbers around groups and single galaxies. Although, the range of physical distances probed around the single and group galaxies is quite different, making a direct comparison challenging.

\section{Discussion and Summary}
\label{sec_discussion}

Spectra of multiple images of strongly-lensed quasars offer the opportunity to tomographically map the foreground absorbing gas. In this work, we leverage the recent MUSE observations of two fields containing multiple images of quasars ($z\approx2.2, 2.8$), that are strongly lensed by galaxy clusters ($z\approx0.5, 0.6$), to study the foreground metal absorption. Thanks to the boost from the magnification effect \citep[median magnification of the multiple quasar images $\approx$2.5--17;][]{Acebron2022a,Acebron2022b}, the MUSE effective integration times on the multiple quasar images are some of the largest, ranging from 30 h to 1400 h, leading to signal-to-noise per pixel of $\approx$30--90 around 7000\,\AA\ in the spectra. The best-fit strong lensing models presented by \citet{Acebron2022a} and \citet{Acebron2022b} are used to accurately estimate the physical transverse separations between the quasar multiple images in the absorber plane. The angular separations between the quasar multiple images range from 2 to 22 arcsec, allowing us to study the low-ionization gas, traced by \mgii\ and \feii\ absorption, over physical separations of 0.4-150\,kpc at $z\approx0.5-2.3$, and the high-ionization gas, traced by \civ\ absorption, over 0.4-25\,kpc at $z\approx1.6-2.3$. 

By looking at how the fractional difference in rest equivalent width ($\Delta W_r$) of the metal absorption between two quasar sightlines varies with the physical separation, we investigate the coherence of metal-enriched gas. The main results of this study are discussed below.  

--- Overall, $\Delta W_r$ shows an increasing trend with physical separation for both the low- and high-ionization gas phase, indicating that the metal absorption becomes less coherent on going to larger separations (Fig.~\ref{fig:diff_sep}). Within a characteristic length scale of $\approx$10 kpc, the metal-enriched gas in the CGM shows a higher level of coherence. However, there is also considerable scatter in $\Delta W_r$, with values ranging from $\approx$0.08 to $\approx$0.9, suggesting clumpy metal-enriched gas in the CGM even at separations smaller than coherent length scale of $\lesssim$10 kpc. Interestingly, we find paired \mgii\ absorption even at $\approx$100-150\,kpc, which could be tracing the same halo. A detailed analysis of the chemical composition and kinematic structure of the absorption lines is required to check whether they trace large-scale spatially-coherent structures.

--- We do not find any significant dependence of $\Delta W_r$ on the equivalent width, redshift, and associated galaxy environment, i.e., single or groups of galaxies (Fig.~\ref{fig:diff_ew} and Fig.~\ref{fig:gal_abs}). 

--- Comparing the coherence of the different metal absorption, we find that \civ\ shows the highest degree of coherence, followed by \mgii\ and \feii, implying that the coherence length of the high-ionization gas phase is likely to be larger than that of the low-ionization gas phase (Fig.~\ref{fig:diff_frac}). At separations less than 10 kpc, about 90, 50, and 30 percent of the \civ, \mgii, and \feii\ absorbers have $\Delta W_r$ less than 0.5, respectively. 

--- Finally, we find cases of coherent metal absorption detected across three quasar sightlines, with the coherence being the highest across the closest quasar sightlines at separations $\lesssim$10 kpc (Fig.~\ref{fig:triple_abs}).

Our results are qualitatively similar to what has been found in previous observations of metal absorption towards lensed quasars \citep[e.g.][]{Ellison2004,Rubin2018,Augustin2021}. Based on a compilation of paired \mgii\ and \civ\ absorption towards lensed quasars in the literature, \citet{Rubin2018} reported similar increase in $\Delta W_r$ with physical separation, as well as significant scatter in the relation, similar to what we find in this work. Consistent with our results, they find that \civ\ systems exhibit a higher degree of coherence than \mgii, with $\approx$90 percent of \civ\ systems having $\Delta W_r$ less than 0.5 at separations less than 10 kpc compared to $\approx$64 percent for \mgii. Using observations of the MUSE Ultra Deep Field, which consists of a bright quasar pair at $z\approx3$, \citet{Fossati2019} detected a system of correlated \mgii\ absorption at $z\approx0.88$ along two sightlines separated by $\approx$480 kpc that has $\Delta W_r$ $\approx$0.68 and is associated with a group of six galaxies. This is in line with the trends we find in this work, and supports the existence of at least some large-scale coherent structures in the cool gas.

In addition to the observational studies, a large spread in $\Delta W_r$ of \mgii\ absorption has been found in high resolution ($\approx$0.1 kpc) FOGGIE simulations of the CGM \citep{Augustin2021}. Further, this study found that extended inflowing filaments show coherence over larger scales than outflowing clumpy structures. Therefore, they suggest that \mgii\ systems that vary strongly over small scales likely trace outflowing clumps rather than filamentary structures. On the other hand, using the Illustris TNG50 simulations, \citet{Nelson2020} found that the CGM is composed of thousands of small ($\approx$1 kpc) cold gas clouds that are predominantly infalling and not outflowing. Additionally, using three dimensional hydrodynamical simulations, \citep{Fielding2020} have found evidence of a fractal nature of the cooling surface that arises within the radiative turbulent mixing layers in a multiphase medium. Further analysis of a statistical sample of galaxy haloes in high-resolution simulations is required to understand the effect of spatial resolution and different feedback models on the origin of cold gas clouds in the CGM \citep[e.g.][]{vandeVoort2019}.

A complimentary and indirect observational technique to map the distribution and extent of CGM gas has typically been to study the radial profiles of equivalent width and covering fraction in statistical ensembles of quasar-galaxy pairs. The trend of decreasing coherence of metal absorption with separation that we observe in this study is qualitatively consistent with the general decreasing trend of covering fraction of metals with projected separation from galaxies reported in the literature \citep[e.g.][]{Chen2010,Nielsen2013,Bordoloi2014,Dutta2020,Dutta2021}. It is interesting to note that such studies find a more extended distribution of \civ\ absorption than \mgii\ around galaxies, and increasing ionization state of the gas with increasing distance from galaxies \citep[e.g.][]{Werk2014,Dutta2021}. Taken together with the results from this study, this could point towards a simple picture where the clumpier low-ionization gas is embedded in a more homogeneous and highly ionized gas. Additionally, studies have found signatures of more extended \mgii\ absorption in overdense group-environments \citep[e.g.][]{Bordoloi2011,Nielsen2018,Dutta2020,Dutta2021}. Here we do not find any discernible difference in the coherence and distribution of \mgii\ absorption between single and groups of galaxies, with the caveat that the sample sizes are statistically small and different in properties.

In the future, a more detailed comparison of the column densities and kinematics of individual absorption components using high resolution ($R\gtrsim20,000$) spectra would place tighter constraints on the sizes of individual gas clouds and larger-scale coherent structures, parameters that are key to constrain models of metal enrichment and mixing. Combining this with the physical properties of the galaxies obtained from the MUSE data, we can study in more detail the connection between structures of the metal-enriched CGM and the galaxy environment.

\section*{Acknowledgements}

We thank the anonymous reviewer for their helpful comments.
This work is based on observations collected at the European Organisation for Astronomical Research in the Southern Hemisphere under ESO programme IDs 0102.A-0642(A), 0103.A-0554(A), 092.B-0512(A).
We acknowledge financial support through grants PRIN-MIUR 2017WSCC32 and 2020SKSTHZ.
AA has received funding from the European Union’s Horizon 2020 research and innovation programme under the Marie Skłodowska-Curie grant agreement No 101024195 - ROSEAU. 
This project has received funding from the European Research Council (ERC) under the European Union's Horizon 2020 research and innovation programme (grant agreement No 757535) and by Fondazione Cariplo (grant No 2018-2329).


\section*{Data Availability}

The observational data used in this work are available from the European Southern Observatory archive: \url{https://archive.eso.org}.



\bibliographystyle{mnras}
\bibliography{paper} 




\appendix

\section{Metal absorption lines detected towards the quasar multiple images}

\begin{table}
\caption{List of \mgii\ and \civ\ systems detected towards image A of J1029.}
\centering 
\begin{tabular}{ccccc}
\hline
Ion & Rest             & Redshift & Equivalent         & Error in $W_r$ \\
    & Wavelength (\AA) &          & Width ($W_r$; \AA) & (\AA)          \\ 
\hline
MgII & 2796 & 0.5111 & 0.832 & 0.011 \\
MgII & 2803 &        & 0.635 & 0.012 \\
MgI  & 2852 &        & 0.114 & 0.012 \\
FeII & 2600 &        & 0.427 & 0.008 \\
MgII & 2796 & 0.9186 & 1.669 & 0.008 \\
MgII & 2803 &        & 1.461 & 0.009 \\
MgI  & 2852 &        & 0.395 & 0.010 \\
FeII & 2600 &        & 1.071 & 0.007 \\
MgII & 2796 & 1.0059 & 0.212 & 0.030 \\
MgII & 2803 &        & 0.094 & 0.017 \\
MgII & 2796 & 1.6931 & 0.988 & 0.021 \\
MgII & 2803 &        & 0.715 & 0.021 \\
MgI  & 2852 &        & 0.064 & 0.020 \\
FeII & 2600 &        & 0.455 & 0.014 \\
MgII & 2796 & 1.8912 & 0.150 & 0.016 \\
MgII & 2803 &        & 0.083 & 0.015 \\
MgII & 2796 & 1.9117 & 1.997 & 0.019 \\
MgII & 2803 &        & 1.749 & 0.019 \\
MgI  & 2852 &        & 0.186 & 0.019 \\
FeII & 2600 &        & 0.891 & 0.024 \\
MgII & 2796 & 1.9417 & 0.650 & 0.019 \\
MgII & 2803 &        & 0.402 & 0.015 \\
MgI  & 2852 &        & 0.069 & 0.014 \\
FeII & 2600 &        & 0.394 & 0.017 \\
CIV  & 1548 & 1.6150 & 0.341 & 0.005 \\
CIV  & 1550 &        & 0.324 & 0.006 \\
SiIV & 1393 &        & 0.105 & 0.006 \\
SiIV & 1402 &        & 0.074 & 0.006 \\
CIV  & 1548 & 1.6228 & 0.338 & 0.005 \\
CIV  & 1550 &        & 0.185 & 0.005 \\
CIV  & 1548 & 1.6930 & 0.327 & 0.008 \\
CIV  & 1550 &        & 0.155 & 0.009 \\
CIV  & 1548 & 1.7210 & 0.102 & 0.005 \\
CIV  & 1550 &        & 0.047 & 0.005 \\
CIV  & 1548 & 1.7651 & 0.043 & 0.003 \\
CIV  & 1550 &        & 0.031 & 0.003 \\
CIV  & 1548 & 1.8908 & 0.505 & 0.009 \\
CIV  & 1550 &        & 0.466 & 0.008 \\
SiIV & 1393 &        & 0.182 & 0.003 \\
SiIV & 1402 &        & 0.272 & 0.003 \\
CIV  & 1548 & 1.8947 & 0.404 & 0.005 \\
CIV  & 1550 &        & 0.333 & 0.005 \\
CIV  & 1548 & 1.8981 & 0.201 & 0.004 \\
CIV  & 1550 &        & 0.152 & 0.006 \\
SiIV & 1393 &        & 0.020 & 0.003 \\
SiIV & 1402 &        & 0.011 & 0.003 \\
CIV  & 1548 & 1.9117 & 1.184 & 0.014 \\
CIV  & 1550 &        & 0.884 & 0.015 \\
SiIV & 1393 &        & 0.882 & 0.008 \\
SiIV & 1402 &        & 0.492 & 0.008 \\
CIV  & 1548 & 2.1083 & 0.133 & 0.003 \\
CIV  & 1550 &        & 0.080 & 0.003 \\
SiIV & 1393 &        & 0.025 & 0.003 \\
SiIV & 1402 &        & 0.013 & 0.003 \\
CIV  & 1548 & 2.1195 & 0.062 & 0.003 \\
CIV  & 1550 &        & 0.030 & 0.003 \\
CIV  & 1548 & 2.1269 & 0.146 & 0.002 \\
CIV  & 1550 &        & 0.108 & 0.002 \\
CIV  & 1548 & 2.1349 & 0.081 & 0.003 \\
CIV  & 1550 &        & 0.043 & 0.003 \\
CIV  & 1548 & 2.1819 & 0.432 & 0.004 \\
CIV  & 1550 &        & 0.301 & 0.004 \\
SiIV & 1393 &        & 0.112 & 0.007 \\
SiIV & 1402 &        & 0.059 & 0.006 \\
CIV  & 1548 & 2.1954 & 1.712 & 0.004 \\
CIV  & 1550 &        & 1.313 & 0.004 \\
\hline
\end{tabular}
\label{tab:j1029a_abs_list}
\end{table}

\begin{table}
\caption{List of \mgii\ and \civ\ systems detected towards image B of J1029.}
\centering 
\begin{tabular}{ccccc}
\hline
Ion & Rest             & Redshift & Equivalent         & Error in $W_r$ \\
    & Wavelength (\AA) &          & Width ($W_r$; \AA) & (\AA)      \\ 
\hline
MgII & 2796 & 0.5124 & 0.109 & 0.006 \\
MgII & 2803 &        & 0.053 & 0.006 \\
FeII & 2600 &        & 0.044 & 0.004 \\
MgII & 2796 & 0.5654 & 0.088 & 0.009 \\
MgII & 2803 &        & 0.038 & 0.008 \\
MgII & 2796 & 0.6731 & 1.729 & 0.010 \\
MgII & 2803 &        & 1.585 & 0.010 \\
MgI  & 2852 &        & 0.612 & 0.010 \\
FeII & 2600 &        & 1.352 & 0.011 \\
MgII & 2796 & 0.9184 & 0.173 & 0.005 \\
MgII & 2803 &        & 0.103 & 0.006 \\
FeII & 2600 &        & 0.004 & 0.003 \\
MgII & 2796 & 1.0032 & 0.613 & 0.020 \\
MgII & 2803 &        & 0.441 & 0.020 \\
MgI  & 2852 &        & 0.044 & 0.006 \\
FeII & 2600 &        & 0.307 & 0.004 \\
MgII & 2796 & 1.7628 & 0.404 & 0.014 \\
MgII & 2803 &        & 0.412 & 0.012 \\
MgI  & 2852 &        & 0.055 & 0.010 \\
MgII & 2796 & 1.8939 & 0.246 & 0.014 \\
MgII & 2803 &        & 0.427 & 0.013 \\
MgII & 2796 & 1.9122 & 1.337 & 0.020 \\
MgII & 2803 &        & 0.926 & 0.020 \\
MgI  & 2852 &        & 0.051 & 0.013 \\
FeII & 2600 &        & 0.342 & 0.019 \\
MgII & 2796 & 1.9788 & 0.401 & 0.019 \\
MgII & 2803 &        & 0.299 & 0.018 \\
MgI  & 2852 &        & 0.035 & 0.013 \\
FeII & 2600 &        & 0.382 & 0.011 \\
CIV  & 1548 & 1.6085 & 0.040 & 0.003 \\
CIV  & 1550 &        & 0.028 & 0.003 \\
CIV  & 1548 & 1.6150 & 0.115 & 0.005 \\
CIV  & 1550 &        & 0.138 & 0.007 \\
CIV  & 1548 & 1.6221 & 0.300 & 0.003 \\
CIV  & 1550 &        & 0.158 & 0.004 \\
CIV  & 1548 & 1.6253 & 0.024 & 0.003 \\
CIV  & 1550 &        & 0.012 & 0.003 \\
CIV  & 1548 & 1.6556 & 0.172 & 0.006 \\
CIV  & 1550 &        & 0.093 & 0.006 \\
CIV  & 1548 & 1.6924 & 0.154 & 0.007 \\
CIV  & 1550 &        & 0.057 & 0.004 \\
CIV  & 1548 & 1.7065 & 0.050 & 0.003 \\
CIV  & 1550 &        & 0.025 & 0.003 \\
CIV  & 1548 & 1.7211 & 0.039 & 0.004 \\
CIV  & 1550 &        & 0.020 & 0.004 \\
CIV  & 1548 & 1.7626 & 0.194 & 0.009 \\
CIV  & 1550 &        & 0.112 & 0.009 \\
CIV  & 1548 & 1.8910 & 0.599 & 0.010 \\
CIV  & 1550 &        & 0.683 & 0.008 \\
SiIV & 1393 &        & 0.138 & 0.005 \\
SiIV & 1402 &        & 0.134 & 0.005 \\
CIV  & 1548 & 1.8943 & 0.707 & 0.005 \\
CIV  & 1550 &        & 0.689 & 0.007 \\
SiIV & 1393 &        & 0.468 & 0.009 \\
SiIV & 1402 &        & 0.425 & 0.009 \\
CIV  & 1548 & 1.9017 & 0.559 & 0.007 \\
CIV  & 1550 &        & 0.387 & 0.008 \\
SiIV & 1393 &        & 0.200 & 0.005 \\
SiIV & 1402 &        & 0.159 & 0.005 \\
CIV  & 1548 & 1.9120 & 0.902 & 0.010 \\
CIV  & 1550 &        & 0.719 & 0.011 \\
SiIV & 1393 &        & 0.688 & 0.006 \\
SiIV & 1402 &        & 0.346 & 0.006 \\
CIV  & 1548 & 2.1083 & 0.200 & 0.005 \\
CIV  & 1550 &        & 0.106 & 0.005 \\
SiIV & 1393 &        & 0.036 & 0.003 \\
SiIV & 1402 &        & 0.051 & 0.004 \\
\hline
\end{tabular}
\label{tab:j1029b_abs_list1}
\end{table}

\begin{table}
\caption{List of \mgii\ and \civ\ systems detected towards image B of J1029 (continued from Table~\ref{tab:j1029b_abs_list1}).}
\centering 
\begin{tabular}{ccccc}
\hline
Ion & Rest             & Redshift & Equivalent         & Error in $W_r$ \\
    & Wavelength (\AA) &          & Width ($W_r$; \AA) & (\AA)      \\     
\hline
CIV  & 1548 & 2.1209 & 0.571 & 0.004 \\
CIV  & 1550 &        & 0.504 & 0.004 \\
MgII & 2796 &        & 0.046 & 0.012 \\
MgII & 2803 &        & 0.025 & 0.011 \\
CIV  & 1548 & 2.1285 & 0.240 & 0.004 \\
CIV  & 1550 &        & 0.182 & 0.004 \\
SiIV & 1393 &        & 0.050 & 0.004 \\
SiIV & 1402 &        & 0.021 & 0.004 \\
CIV  & 1548 & 2.1349 & 0.071 & 0.003 \\
CIV  & 1550 &        & 0.031 & 0.003 \\
CIV  & 1548 & 2.1820 & 0.386 & 0.004 \\
CIV  & 1550 &        & 0.234 & 0.004 \\
SiIV & 1393 &        & 0.144 & 0.006 \\
SiIV & 1402 &        & 0.093 & 0.006 \\
CIV  & 1548 & 2.1955 & 1.579 & 0.004 \\
CIV  & 1550 &        & 1.236 & 0.004 \\
\hline
\end{tabular}
\label{tab:j1029b_abs_list2}
\end{table}

\begin{table}
\caption{List of \mgii\ and \civ\ systems detected towards image C of J1029.}
\centering 
\begin{tabular}{ccccc}
\hline 
Ion & Rest             & Redshift & Equivalent         & Error in $W_r$ \\
    & Wavelength (\AA) &          & Width ($W_r$; \AA) & (\AA)          \\ 
\hline
MgII & 2796 & 0.9182 & 0.596 & 0.028 \\
MgII & 2803 &        & 0.267 & 0.022 \\
MgII & 2796 & 1.0032 & 0.566 & 0.025 \\
MgII & 2803 &        & 0.425 & 0.025 \\
MgI  & 2852 &        & 0.140 & 0.029 \\
FeII & 2600 &        & 0.196 & 0.023 \\
MgII & 2796 & 1.8939 & 0.227 & 0.016 \\
MgII & 2803 &        & 0.221 & 0.015 \\
MgII & 2796 & 1.9122 & 0.712 & 0.022 \\
MgII & 2803 &        & 0.521 & 0.023 \\
FeII & 2600 &        & 0.136 & 0.022 \\
MgII & 2796 & 1.9417 & 0.264 & 0.021 \\
MgII & 2803 &        & 0.085 & 0.016 \\
MgII & 2796 & 1.9788 & 0.344 & 0.020 \\
MgII & 2803 &        & 0.277 & 0.030 \\
MgI  & 2852 &        & 0.097 & 0.022 \\
FeII & 2600 &        & 0.196 & 0.015 \\
CIV  & 1548 & 2.1077 & 0.278 & 0.020 \\
CIV  & 1550 &        & 0.156 & 0.016 \\
CIV  & 1548 & 2.1206 & 0.652 & 0.014 \\
CIV  & 1550 &        & 0.688 & 0.017 \\
MgII & 2796 &        & 0.117 & 0.018 \\
MgII & 2803 &        & 0.082 & 0.015 \\
CIV  & 1548 & 2.1957 & 1.576 & 0.007 \\
CIV  & 1550 &        & 1.166 & 0.007 \\
\hline
\end{tabular}
\label{tab:j1029c_abs_list}
\end{table}

\begin{table}
\caption{List of \mgii\ and \civ\ systems detected towards image A of J2222.}
\centering 
\begin{tabular}{ccccc}
\hline
Ion & Rest             & Redshift & Equivalent         & Error in $W_r$ \\
    & Wavelength (\AA) &          & Width ($W_r$; \AA) & (\AA)          \\ 
\hline
MgII & 2796 & 1.2017 & 1.191 & 0.030 \\
MgII & 2803 &        & 1.154 & 0.036 \\
MgI  & 2852 &        & 0.410 & 0.029 \\
FeII & 2600 &        & 0.861 & 0.033 \\
MgII & 2796 & 2.2981 & 1.016 & 0.034 \\
MgII & 2803 &        & 1.110 & 0.029 \\
FeII & 2600 &        & 0.391 & 0.023 \\
CIV  & 1548 &        & 1.037 & 0.023 \\
CIV  & 1550 &        & 1.081 & 0.020 \\
CIV  & 1548 & 2.8022 & 1.155 & 0.014 \\
CIV  & 1550 &        & 0.912 & 0.011 \\
SiIV & 1393 &        & 0.503 & 0.020 \\
SiIV & 1402 &        & 0.230 & 0.015 \\
\hline
\end{tabular}
\label{tab:j2222a_abs_list}
\end{table}

\begin{table}
\caption{List of \mgii\ and \civ\ systems detected towards image B of J2222.}
\centering 
\begin{tabular}{ccccc}
\hline
Ion & Rest             & Redshift & Equivalent         & Error in $W_r$ \\
    & Wavelength (\AA) &          & Width ($W_r$; \AA) & (\AA)          \\ 
\hline
MgII & 2796 & 1.2017 & 1.739 & 0.030 \\
MgII & 2803 &        & 1.556 & 0.026 \\
MgI  & 2852 &        & 0.597 & 0.032 \\
FeII & 2600 &        & 1.115 & 0.036 \\
MgII & 2796 & 2.2985 & 0.917 & 0.032 \\
MgII & 2803 &        & 1.105 & 0.031 \\
FeII & 2600 &        & 0.394 & 0.024 \\
CIV  & 1548 &        & 1.158 & 0.023 \\
CIV  & 1550 &        & 1.608 & 0.023 \\
CIV  & 1548 & 2.8022 & 1.102 & 0.012 \\
CIV  & 1550 &        & 0.904 & 0.014 \\
SiIV & 1393 &        & 0.482 & 0.019 \\
SiIV & 1402 &        & 0.257 & 0.017 \\
\hline
\end{tabular}
\label{tab:j2222b_abs_list}
\end{table}

\begin{table}
\caption{List of \mgii\ and \civ\ systems detected towards image C of J2222.}
\centering 
\begin{tabular}{ccccc}
\hline
Ion & Rest             & Redshift & Equivalent         & Error in $W_r$ \\
    & Wavelength (\AA) &          & Width ($W_r$; \AA) & (\AA)          \\  
\hline
MgII & 2796 & 1.2965 & 0.805 & 0.041 \\
MgII & 2803 &        & 0.622 & 0.043 \\
MgI  & 2852 &        & 0.291 & 0.044 \\
FeII & 2600 &        & 0.382 & 0.029 \\
MgII & 2796 & 2.2990 & 1.819 & 0.055 \\
MgII & 2803 &        & 1.856 & 0.050 \\
FeII & 2600 &        & 0.186 & 0.045 \\
CIV  & 1548 &        & 1.655 & 0.038 \\
CIV  & 1550 &        & 1.677 & 0.035 \\
CIV  & 1548 & 2.8022 & 1.082 & 0.012 \\
CIV  & 1550 &        & 0.858 & 0.014 \\
SiIV & 1393 &        & 0.507 & 0.031 \\
SiIV & 1402 &        & 0.191 & 0.025 \\
\hline
\end{tabular}
\label{tab:j2222c_abs_list}
\end{table}


\bsp	
\label{lastpage}
\end{document}